\documentclass[12pt]{article}

\usepackage[margin=1in]{geometry}
\usepackage{times}
\usepackage{enumerate}
\usepackage{hyperref}
\usepackage{subfiles}
\usepackage{url}
\usepackage{xcolor}
\usepackage{xspace}
\usepackage{balance}
\usepackage[noadjust]{cite}
\usepackage{siunitx}

\usepackage{amsmath}
\usepackage{amssymb}
\usepackage{amsbsy}
\usepackage{bm}

\usepackage{algorithm}
\usepackage{algorithmic}
\usepackage{graphicx}
\usepackage{float}
\usepackage{subfigure}
\usepackage{tabu}
\usepackage{multirow}
\usepackage[export]{adjustbox}
\graphicspath{{./fig/}}

\usepackage{tikz}
\usepackage{varwidth}

\newcommand{\citep}[1]{\cite{#1}}

\def \Cramer{Cram\'{e}r}

\def \ie{\emph{i.e.}\xspace}
\def \eg{\emph{e.g.}\xspace}
\def \cf{\emph{cf.}\xspace}
\def \invivo{in vivo\xspace}

\def \exvivo{ex vivo\xspace}
\def \Invivo{In vivo\xspace}
\def \Invitro{In vitro\xspace}

\def \Exvivo{Ex vivo\xspace}

\let \oldfootnote \footnote
\def \footnote{\ifhmode\unskip\fi\oldfootnote}

\def \regis{\textsuperscript{\textregistered}\xspace}
\def \tmark{\textsuperscript{\texttrademark}\xspace}
\def \matlab{MATLAB\regis}

\definecolor{mat-red}{rgb}{0.6350, 0.0780, 0.1840}
\definecolor{mat-orange}{rgb}{0.8500, 0.3250, 0.0980}
\definecolor{mat-yellow}{rgb}{0.9290, 0.6940, 0.1250}
\definecolor{mat-purple}{rgb}{0.4940, 0.1840, 0.5560}
\definecolor{mat-green}{rgb}{0.4660, 0.6740, 0.1880}
\definecolor{mat-cyan}{rgb}{0.3010, 0.7450, 0.9330}

\newcommand{\RA}{\textcolor{mat-red}{AR}\xspace}
\newcommand{\LA}{\textcolor{mat-orange}{AL}\xspace}
\newcommand{\RP}{\textcolor{mat-yellow}{PR}\xspace}
\newcommand{\LP}{\textcolor{mat-purple}{PL}\xspace}
\newcommand{\IC}{\textcolor{mat-green}{IC}\xspace}
\newcommand{\AC}{\textcolor{mat-cyan}{AC}\xspace}

\newcommand{\WM}{\textcolor{mat-red}{WM}\xspace}
\newcommand{\GM}{\textcolor{mat-cyan}{GM}\xspace}

\let \originalleft \left
\let \originalright \right
\renewcommand{\left}{\mathopen{}\mathclose\bgroup\originalleft}
\renewcommand{\right}{\aftergroup\egroup\originalright}

\newcommand{\brac}[1]{\left[#1\right]}
\newcommand{\set}[1]{\left\{#1\right\}}
\newcommand{\abs}[1]{\left\lvert #1 \right\rvert}
\newcommand{\paren}[1]{\left(#1\right)}
\newcommand{\norm}[1]{\left\|#1\right\|}

\newcommand{\argmin}[1]{\operatorname{arg}\,\min_{#1}\,}

\newcommand{\where}{\,\, \mathrm{where}}

\newcommand{\del}{\partial}
\newcommand{\dela}[1]{\frac{\del}{\del#1}\xspace}
\newcommand{\grad}{\nabla}
\newcommand{\grada}[1]{\grad_{#1}}

\newcommand{\est}[1]{\widehat{#1}}
\newcommand{\esta}[2]{\widehat{#1}\paren{#2}}

\newcommand{\real}{\mathbb{R}}
\newcommand{\reals}[1]{\real^{#1}}
\newcommand{\complex}{\mathbb{C}}
\newcommand{\complexes}[1]{\complex^{#1}}

\newcommand{\Ltwo}{\mathcal{L}^2}

\newcommand{\gauss}[2]{\mathcal{N}\paren{#1,#2}}
\newcommand{\cgauss}[2]{\mathbb{C}\gauss{#1}{#2}}
\newcommand{\unif}[1]{\operatorname{unif}\paren{#1}}
\newcommand{\logunif}[1]{\operatorname{logunif}\paren{#1}}

\newcommand{\tpose}{^{\mathsf{T}}}
\newcommand{\ctpose}{^{\mathsf{H}}}

\newcommand{\inv}[1]{\paren{#1}^{-1}}

\newcommand{\eye}[1]{\mathbf{I}_{#1}}
\newcommand{\ones}[1]{\boldsymbol{1}_{#1}}
\newcommand{\zeros}[1]{\boldsymbol{0}_{#1}}
\newcommand{\diag}[1]{\operatorname{diag}\paren{#1}}
\newcommand{\trace}[1]{\operatorname{tr}\paren{#1}}

\newcommand{\expect}[2]{\mathsf{E}_{#1}\paren{#2}}

\newcommand{\dist}[1]{\mathsf{p}_{#1}}

\newcommand{\degrees}[1]{#1^{\circ}}

\newcommand{\expa}[1]{\exp{\paren{#1}}}

\newcommand{\snr}[1]{\mathsf{SNR}\paren{#1}}

\newcommand{\mnstd}[2]{$#1\pm#2$}

\newcommand{\To}{T_1}
\newcommand{\Tt}{T_2}
\newcommand{\TR}{T_\mathrm{R}}

\newcommand{\fast}{\mathrm{F}}
\newcommand{\slow}{\mathrm{S}}
\newcommand{\tf}[1]{T_{#1,\fast}}
\newcommand{\ts}[1]{T_{#1,\slow}}
\newcommand{\TE}{T_\mathrm{E}}
\newcommand{\dess}{s_{\mathrm{D}}}
\newcommand{\ff}{f_\fast}
\newcommand{\mzero}{m_0}
\newcommand{\Rtpf}{R'_{2,\fast}}
\newcommand{\Rtps}{R'_{2,\slow}}
\newcommand{\ompfmed}{\omega_{\fast}}
\newcommand{\ompsmed}{\omega_{\slow}}
\newcommand{\flip}{\alpha}

\newcommand{\bmy}{\mathbf{y}}
\newcommand{\bms}{\mathbf{s}}
\newcommand{\bmsa}[1]{\bms\paren{#1}}
\newcommand{\bmx}{\mathbf{x}}
\newcommand{\bmnu}{\boldsymbol{\nu}}
\newcommand{\bmP}{\mathbf{P}}
\newcommand{\bmeps}{\boldsymbol{\epsilon}}
\newcommand{\bmSig}{\boldsymbol{\Sigma}}
\newcommand{\bmF}{\mathbf{F}}
\newcommand{\bmFa}[1]{\bmF\paren{#1}}
\newcommand{\cost}{\Psi}
\newcommand{\costa}[1]{\Psi\paren{#1}}
\newcommand{\bmW}{\mathbf{W}}
\newcommand{\bmPstar}{\bmP^*}
\newcommand{\setP}{\mathbb{P}}
\newcommand{\expcost}{\bar{\cost}}
\newcommand{\bmp}{\mathbf{p}}

\newcommand{\bmq}{\mathbf{q}}
\newcommand{\bma}{\mathbf{a}}
\newcommand{\bmb}{\mathbf{b}}
\newcommand{\bmX}{\mathbf{X}}
\newcommand{\bmM}{\mathbf{M}}
\newcommand{\bmK}{\mathbf{K}}
\newcommand{\bmk}{\mathbf{k}}
\newcommand{\bmka}[1]{\bmk\paren{#1}}
\newcommand{\bmL}{\boldsymbol{\Lambda}}
\newcommand{\bmG}{\boldsymbol{\Gamma}}
\newcommand{\bmz}{\mathbf{z}}
\newcommand{\bmza}[1]{\bmz\paren{#1}}
\newcommand{\bmZ}{\mathbf{Z}}
\newcommand{\bmm}{\mathbf{m}}
\newcommand{\bmmx}{\bmm_{\bmx}}
\newcommand{\bmmz}{\bmm_{\bmz}}
\newcommand{\bmC}{\mathbf{C}}
\newcommand{\bmCxz}{\bmC_{\bmx\bmz}}
\newcommand{\bmCzz}{\bmC_{\bmz\bmz}}

\newcommand{\mwf}{f_{\mathrm{M}}}
\newcommand{\flipnom}{\flip_0}
\newcommand{\stx}{\kappa}
\newcommand{\TRd}{T_{\mathrm{R},d}}
\newcommand{\bmmymag}{\mathbf{m}_{\abs{\bmy}}}
\newcommand{\bmmnu}{\mathbf{m}_{\bmnu}}
\newcommand{\bmA}{\mathbf{A}}
\newcommand{\setX}{\mathcal{X}}
\newcommand{\mwfest}{\est{f}_{\mathrm{M}}}
\newcommand{\bmi}{\mathbf{i}}
\newcommand{\bmyt}{\widetilde{\bmy}}
\newcommand{\bmepst}{\widetilde{\bmeps}}

\newcommand{\ffest}{\est{f}_{\mathrm{F}}}

\usetikzlibrary{shapes}
\usetikzlibrary{arrows}
\usetikzlibrary{intersections}
\usepackage{tkz-euclide}
\usetkzobj{all}

\tikzstyle{input} = [
  draw=black,
  trapezium,
  trapezium left angle=75,
  trapezium right angle=105,
  fill=green!20,
  text centered,
  minimum height=2em
]

\tikzstyle{block} = [
  rectangle,
  draw=black,
  fill=blue!20,
  rounded corners,
  text centered,
  minimum height=2em
]

\tikzstyle{sum} = [
  circle,
  draw=black,
  radius=0.2
]

\tikzstyle{decision} = [
  diamond,
  draw=black,
  fill=yellow!20,
  text badly centered,
  text width=5em
]

\tikzstyle{output} = [
  draw=black,
  ellipse,
  fill=red!20,
  text centered,
  minimum height=2em
]

\tikzstyle{arrow} = [
  ->, 
  very thick
]

\tikzstyle{line} = [
  -,
  very thick
]

\title{%
	Fast, Precise Myelin Water Quantification \\
	using DESS MRI and Kernel Learning
}%

\author{%
	Gopal~Nataraj$^\star$, %
	Jon-Fredrik~Nielsen$^\dagger$, %
	Mingjie~Gao$^\star$, %
	and %
	Jeffrey~A.~Fessler$^\star$
}

\date{%
	$^\star$Dept. of Electrical Engineering and Computer Science, 
	University of Michigan \\
	$^\dagger$Dept. of Biomedical Engineering, University of Michigan \\
}

\begin{document}
\maketitle

\begin{abstract}
\label{f,abs}

\textbf{Purpose:}
To investigate the feasibility 
of myelin water content quantification
using fast dual-echo steady-state (DESS) scans
and machine learning with kernels.

\textbf{Methods:}
We optimized combinations of steady-state (SS) scans
for precisely estimating the fast-relaxing signal fraction $\ff$ 
of a two-compartment signal model,
subject to a scan time constraint.
We estimated $\ff$ 
from the optimized DESS acquisition
using a recently developed method
for rapid parameter estimation 
via regression with kernels (PERK).
We compared DESS PERK $\ff$ estimates 
to conventional myelin water fraction (MWF) estimates
from a longer multi-echo spin-echo (MESE) acquisition
in simulation, \invivo, and \exvivo studies. 

\textbf{Results:}
Simulations demonstrate 
that DESS PERK $\ff$ estimators
and MESE MWF estimators
achieve comparable error levels.
\Invivo and \exvivo experiments demonstrate
that MESE MWF and DESS PERK $\ff$ estimates
are quantitatively comparable measures
of WM myelin water content.
To our knowledge,
these experiments are the first
to demonstrate myelin water images 
from a SS acquisition
that are quantitatively similar
to conventional MESE MWF images.

\textbf{Conclusion:}
Combinations of fast DESS scans
can be designed 
to enable precise $\ff$ estimation.
PERK is well-suited for $\ff$ estimation.
DESS PERK $\ff$ and MESE MWF estimates
are quantitatively similar measures
of WM myelin water content.


\textbf{Funding Information:}
National Institutes of Health grant P30 AG053760,
University of Michigan MCubed seed grant,
University of Michigan predoctoral fellowship

\end{abstract}

\vspace{3cm}
\begin{center}
	\large{Submitted to Magnetic Resonance in Medicine}
\end{center}
\newpage

\section{Introduction}
\label{m,intro}

Myelin is a lipid-rich material
that forms an insulating sheath
encasing neuronal axons 
predominantly in white matter (WM) regions
of the human brain
\citep{morell:84}.
Demyelination (\ie, myelin loss) is central
to the development 
of several neurodegenerative disorders
such as multiple sclerosis (MS)
\citep{goldenberg:12:msr}. 
Non-invasive myelin quantification in WM
is thus desirable 
for monitoring the onset and progression
of neurodegenerative disease.

MR relaxation time constants
(especially spin-spin time constant $\Tt$)
depend on the macromolecular environment
surrounding water molecules.
In nerve tissue,
these environments vary spatially
on scales much smaller 
than the millimeter-scale resolutions
used in typical MR imaging experiments.
Many researchers have attempted 
to characterize nerve tissue microstructure
by estimating the intravoxel distribution
of MR relaxation time constants
and associating certain ranges of time constants
with particular ``compartments'' or ``pools'' of water molecules
that exist in similar macromolecular environments.
\Invitro NMR studies
of nerve animal tissue
ascribed a fast-relaxing water compartment
with $\Tt\sim$10-40ms 
initially to general protein 
and phospholipid structures \citep{vasilescu:78:wci}
and later more specifically
to water trapped between
the phospholipid bilayers 
of myelin
\citep{menon:91:aoc, stewart:93:ssr}.
Shortly thereafter,
the first MR images 
of so-called \emph{myelin water fraction} (MWF),
defined as the proportion of MR signal 
arising from the fast-relaxing water compartment
relative to total MR signal,
were demonstrated \invivo
in the human brain \citep{mackay:94:ivv}.
More recently,
MWF has been shown 
to correlate well 
with histological measurements
of myelin content 
in animal models
of nerve injury \citep{gareau:00:mta}
and demyelination \citep{webb:03:imt}.
In humans, 
MWF has been measured 
to be significantly lower
in ``normally appearing'' WM 
of MS patients versus controls \citep{laule:04:wca},
and to correlate strongly 
with post-mortem histological measurements
of myelin content
in MS patients \citep{laule:06:mwi}.
These studies provide growing evidence
that MWF as defined in \citep{mackay:94:ivv}
is a specific quantitative marker
of intact WM myelin content.

All of the aforementioned studies
estimate MWF images
from a multi-echo spin echo (MESE) MRI pulse sequence
\citep{carr:54:eod}
with long repetition time $\TR\geq2$s
to ensure sufficient recovery
of the longitudinal magnetization 
in nerve tissue.
Whole-brain MWF imaging 
using such long-$\TR$ MESE acquisitions
at a typical imaging resolution
would require hours of scan time.
To enable more clinically practical scan times,
researchers have more recently shown
that MESE-based MWF imaging
can be accelerated
without significantly changing
the resulting MWF images \citep{does:00:rat,prasloski:12:rwc}
by acquiring multiple gradient echoes
per refocusing pulse \citep{feinberg:91:gga}.
However, 
these and other acquisition modifications
used in \citep{prasloski:12:rwc}
do not address the fundamental long-$\TR$ requirement
of MESE acquisitions
and thus would still require long scan times
for whole-brain MWF imaging
at millimeter-scale resolution.
Furthermore,
estimating a $\Tt$ distribution
from MESE data
constitutes a poorly conditioned estimation problem
that continues to demand high SNR 
\citep{alonsoortiz:15:mbm,does::ibt},
so the need remains 
for a more SNR-efficient acquisition
for myelin water imaging.
As an alternative to MESE acquisitions,
scan profiles consisting of short-$\TR$ steady-state (SS) sequences
were proposed 
for whole-brain myelin water imaging in about 30m scan time
\citep{deoni:08:gmt}.
Despite more recent further refinements
\citep{deoni:11:com, deoni:13:oct},
myelin water images from SS pulse sequences 
have thus far been shown
to be incomparable with MWF images
from MESE pulse sequences
\citep{zhang:15:com},
likely due at least in part
to insufficient precision \citep{lankford:13:oti}
for reasonable scan times.

Inspired by \citep{nataraj:17:oms},
we reconsidered myelin water imaging from SS pulse sequences
from the perspective of statistical experiment design.
In \citep{nataraj:17:oms},
we optimized several combinations
of spoiled gradient-recalled echo (SPGR)
\citep{zur:91:sot}
and dual-echo steady-state (DESS) scans
\citep{redpath:88:fan, bruder:88:ans}
for single-compartment $\To,\Tt$ estimation
and found that different optimized scan combinations
produced significantly different \invivo $\Tt$ estimates
(but comparable phantom $\Tt$ estimates),
indicating \invivo sensitivity to model non-idealities.
Further simulation studies 
suggested that these inconsistencies may be attributable
to multi-compartmental relaxation.
This paper demonstrates
that this apparent SPGR/DESS sensitivity
to multi-compartmental relaxation
can be exploited 
for fast, precise myelin water imaging. 

This paper introduces a new method
\footnote{%
	This paper substantially extends 
	our previous work in myelin water imaging.
	Conference paper \citep{nataraj:17:dfm}
	introduced the estimation algorithm used herein
	but presented simulation results only.
	Conference proceeding \citep{nataraj:17:mwf}
	introduced the experimental design algorithm used herein
	but did not compare results
	against conventional MESE MWF estimates.
}
for myelin water content quantification
based on a fast SS MRI acquisition
and PERK \citep{nataraj:18:dfm},
a recently developed learning-inspired algorithm
for fast, scalable MRI parameter estimation.
The acquisition consists
of a combination of DESS scans
optimized to enable precise estimation
of the fast-relaxing signal fraction $\ff$
in two-compartment signal models,
subject to a total scan time constraint.
The PERK estimator learns 
a globally optimal regression function
that nonlinearly maps DESS measurements
to $\ff$ estimates 
using simulated training points,
kernel functions,
and convex optimization.
Our precision-optimized DESS acquisition 
is as fast as the SS acquisition 
proposed in \citep{deoni:11:com}
but enables $\sim$$40$\% expected coefficient of variation
in unbiased $\ff$ estimates.
(Similar calculations for \citep{deoni:08:gmt,deoni:11:com}
found that $\ff$ coefficients of variation
frequently exceeded $100$\% \citep{lankford:13:oti}.)
\Invivo and \exvivo experiments demonstrate
that DESS PERK $\ff$ estimates
and conventional MESE MWF estimates
are quantitatively comparable measures
of WM myelin water content.
To our knowledge,
these experiments are the first 
to demonstrate myelin water images 
from a SS acquisition
that are quantitatively similar
to MESE MWF images.

\section{Theory}
\label{m,theory}

This section highlights several unconventional aspects 
of our myelin water imaging framework.
\S\ref{m,theory,model} describes
a two-compartment DESS signal model.
\S\ref{m,theory,acq} develops 
a scalable method for scan optimization.
\S\ref{m,theory,perk} overviews
Parameter Estimation via Regression with Kernels (PERK) 
\citep{nataraj:17:dfm,nataraj:18:dfm},
a recently developed machine learning algorithm
for fast multiple-parameter estimation.

\subsection{A Two-Compartment DESS Signal Model}
\label{m,theory,model}

We assume that MR signal arises 
from two intra-voxel water compartments:
a fast-relaxing compartment
characterized by comparatively short 
spin-lattice $\tf{1}$ 
and spin-spin $\tf{2}$ relaxation times
and a slow-relaxing compartment
characterized by longer relaxation times $\ts{1},\ts{2}$. 
If these compartments are allowed to exchange,
the resulting DESS signal models 
are difficult to express exactly
\footnote{%
	However,
	we have derived 
	approximate two-compartment DESS models \cite[Ch.~6]{nataraj:18:aiq}
	in the presence of first-order exchange.
}
(unlike analogous SPGR \citep{spencer:00:mos}
or balanced steady-state free precession \citep{deoni:08:iea} models)
due to strongly time-dependent off-resonance effects
imparted by unbalanced DESS dephasing gradients.
We focus here on the non-exchanging case for simplicity
(as is also commonly done
in MESE MWF imaging).

Assuming the DESS echoes are acquired
at symmetric echo times $t \gets \mp\TE$ 
before and after a near-instantaneous RF pulse 
centered at time $t \gets 0$
(where $\gets$ denotes assignment),
we have shown \cite[Ch.~6]{nataraj:18:aiq}
through an analysis similar 
to those in \citep{spencer:00:mos, deoni:08:gmt}
that the non-exchanging noiseless DESS signals can,
to within constants,
be intuitively written as
a sum over compartmental signal contributions:
\begin{align}
	\dess\paren{-\TE} &\propto
		-i\mzero \tan\frac{\flip}{2}
		\label{eq:model,dess-ref} \\
	&\times
		\bigg(
			\ff \paren{1-\eta\paren{\tf{1},\tf{2}}}
			e^{+\paren{1/\tf{2}-\Rtpf+i\ompfmed}\TE}
			\nonumber \\
	&+
			(1-\ff) \paren{1-\eta\paren{\ts{1},\ts{2}}}
			e^{+\paren{1/\ts{2}-\Rtps+i\ompsmed}\TE}
		\bigg);
		\nonumber \\
	\dess\paren{+\TE} &\propto
		+i\mzero \tan{\frac{\flip}{2}} 
		\label{eq:model,dess-def} \\
	&\times 
		\bigg(
			\ff \paren{1-\frac{\eta\paren{\tf{1},\tf{2}}}{\xi\paren{\tf{1}}}}
			e^{-\paren{1/\tf{2}+\Rtpf+i\ompfmed}\TE}
			\nonumber \\
	&+	
			(1-\ff) \paren{1-\frac{\eta\paren{\ts{1},\ts{2}}}{\xi\paren{\ts{1}}}}
			e^{-\paren{1/\ts{2}+\Rtps+i\ompsmed}\TE}
		\bigg),
		\nonumber
\end{align}
where $\eta,\xi$ are intermediate functions defined as
\begin{align}
	\eta\paren{t',t''} &:=
		\sqrt{\frac{1-\paren{\expa{-\TR/t''}}^2}
		{1-\paren{\expa{-\TR/t''}/\xi\paren{t'}}^2}};
		\nonumber \\
	\xi\paren{t'} &:=
		\frac{1-\expa{-\TR/t'}\cos{\flip}}
		{\expa{-\TR/t'}-\cos{\flip}}.
\end{align}
Here, 
$\ff \in \brac{0,1}$ 
denotes fast-relaxing compartmental fraction;
$\mzero$ denotes total spin density; 
$\Rtpf,\Rtps$ 
and $\ompfmed,\ompsmed$ respectively denote 
compartment-specific broadening bandwidths
and median off-resonance frequencies;
$\flip$ denotes flip angle;
and $\TR$ denotes repetition interval.
These expressions assume
that off-resonance distributions 
are independent across compartments,
with Cauchy-distributed marginals.
Though this is perhaps a strong assumption,
it serves to clearly demonstrate
that unlike in the single-compartment case
\cite[Eq.~17]{nataraj:17:oms},
off-resonance effects may not be aggregated 
into an apparent spin density -- 
in fact,
uncompensated off-resonance terms
to first order would influence
the apparent compartmental fractions
of typical interest.

\subsection{A Bayesian Approach to Acquisition Design}
\label{m,theory,acq}

This subsection develops our approach
to designing a fast SS acquisition 
that enables precise $\ff$ estimation. 
For clarity in presentation,
we describe the method here 
in a general manner
and provide implementation details
in \S\ref{m,meth,acq}.
Our method is related 
to the Bayesian \Cramer-Rao Bound \citep{gill:95:aot},
which has been applied previously
in quantitative MRI \citep{akcakaya:15:ots,lewis:16:ddo}
though with a different cost function
than the one developed in this section.
We propose an intuitive cost function
that is amenable to gradient-based optimization
and is thus suitable 
for multi-dimensional parameter estimation problems.

After image reconstruction,
many quantitative MRI acquisitions
produce at each voxel position
a sequence of measurements
$\bmy \in \complexes{D}$,
modeled here as
\begin{align}
	\bmy = \bmsa{\bmx,\bmnu,\bmP} + \bmeps.
	\label{eq:acq,model}
\end{align}
Here, 
$\bms : \reals{L+K+AD} \mapsto \complexes{D}$
models $D$ noiseless signals;
$\bmx \in \reals{L}$ 
denotes $L$ \emph{latent} (\ie, unknown) parameters;
$\bmnu \in \reals{K}$
denotes $K$ \emph{known} parameters;
$\bmP \in \reals{A \times D}$
collects $A$ \emph{acquisition} parameters
for each of $D$ measurements;
and $\bmeps \sim \cgauss{\zeros{D}}{\bmSig}$ 
denotes complex Gaussian noise
with zero mean $\zeros{D} \in \reals{D}$ 
and known covariance $\bmSig \in \reals{D \times D}$.
(As a concrete example,
for single-compartment $\Tt$ estimation 
from spin echo measurements,
$\bmx$ could collect $\mzero,\Tt$;
$\bmnu$ could collect 
known main and RF transmit field inhomogeneities;
and
$\bmP$ could collect $D$ echo times.)
We seek to design $\bmP$ 
to enable precise estimation 
of one or more elements of $\bmx$.

We approach acquisition design
by minimizing a cost function
that characterizes estimation imprecision.
To develop this cost function, 
we utilize the \Cramer-Rao Bound \citep{cramer:46},
which states 
that the covariance of any unbiased estimator of $\bmx$
is bounded below
by the inverse (if it exists)
of the Fisher information matrix
\begin{align}
	\bmFa{\bmx, \bmnu, \bmP} :=
		\paren{\grada{\bmx} \bmsa{\bmx,\bmnu,\bmP}}\ctpose
    \bmSig^{-1} \grada{\bmx} \bmsa{\bmx,\bmnu,\bmP},
  \label{eq:acq,fisher}
\end{align}
where $\grada{\bmx}{}$ 
denotes row gradient with respect to $\bmx$
and $\paren{\cdot}\ctpose$ 
denotes conjugate transpose.
We focus on minimizing a weighted average 
of latent parameter variance lower bounds
\begin{align}
	\costa{\bmx,\bmnu,\bmP} := 
		\trace{\bmW \bmF^{-1}\paren{\bmx,\bmnu,\bmP} \bmW}
	\label{eq:acq,cost}
\end{align}
where $\trace{\cdot}$ 
denotes the matrix trace operation
and $\bmW \in \reals{L \times L}$ 
is a diagonal weighting matrix.
Directly optimizing \eqref{eq:acq,cost} over $\bmP$
would encourage precise estimation
only for some specific $\bmx,\bmnu$ values.
In \citep{nataraj:17:oms},
we addressed this dependence 
of $\cost$ on $\bmx,\bmnu$
through a min-max optimization problem.
The associated ``worst-case'' design criterion
required relatively mild assumptions
on the distribution of $\bmx,\bmnu$
but was non-differentiable in $\bmP$.
For the lower-dimensional application
studied in \citep{nataraj:17:oms},
we optimized the min-max criterion 
through exhaustive grid search,
so non-differentiability did not matter.
However,
grid search scales poorly with $D$
and $D \geq L$ for well-conditioned estimation,
so the need for an alternative approach increases
for higher-$L$ estimation problems.
We study here
an alternate design criterion
that is amenable to gradient-based optimization.
Specifically,
we seek $\bmP$
that minimizes the \emph{expected} weighted average
of latent parameter variance lower bounds
over an acquisition parameter design search space $\setP$:
\begin{align}
	\bmPstar &\in 
		\set{\argmin{\bmP \in \setP} \expcost\paren{\bmP}}, \where
		\label{eq:acq,P-star} \\
	\expcost\paren{\bmP} &:= 
		\expect{\bmx,\bmnu}{\costa{\bmx,\bmnu,\bmP}}
		\label{eq:acq,expcost}
\end{align}
and $\expect{\bmx,\bmnu}{\cdot}$ denotes joint expectation
with respect to a prior joint distribution on $\bmx,\bmnu$.
Under certain conditions
(detailed in the Appendix),
expected cost $\expcost$ is differentiable in $\bmP$ 
and is thus amenable
to gradient-based local optimization.

\subsection{PERK: A Fast Algorithm for Multiple-Parameter Estimation}
\label{m,theory,perk}

This subsection overviews PERK,
a fast machine learning algorithm
for dictionary-free per-voxel MRI parameter estimation.
PERK is a computationally efficient alternative
to conventional dictionary-based grid search estimation:
specifically,
PERK may scale better than grid search
with the number of unknowns
(see \S\ref{s,sim} and \S\ref{s,disc} respectively 
for supporting results and discussion).   
We recently developed and demonstrated PERK 
for single-compartment $\To,\Tt$ estimation
\citep{nataraj:18:dfm}
and a full description is provided therein.	
Here we review the PERK estimator
at a conceptual level;
the Appendix reviews selected mathematical details
and 
\S\ref{s,est} explains our implementation
for myelin water imaging.

In essence,
PERK learns a simple nonlinear estimator
from simulated labeled training points
and evaluates the learned estimator 
on unlabeled test data.
PERK first samples 
prior parameter and noise distributions 
and evaluates signal model \eqref{eq:acq,model} many times
to form a set of parameter-measurement tuples.
The goal of PERK is to then learn
from these labeled training points
a suitable estimator
that maps each measurement and known parameter realization
to a reasonable latent parameter estimate.
This supervised learning problem
is subject to a tradeoff
between expressivity and training complexity:
more complicated estimator architectures 
have the potential 
to accurately represent 
a larger space of functions
but are computationally harder to train.
Towards good expressivity, 
researchers have recently proposed
to learn deep neural network architectures 
for quantitative MRI
\citep{virtue:17:btr,cohen:18:mfd}.
However,
it is well known 
that deep learning requires 
enormous numbers of training points
to train many hyperparameters without overfitting,
and its limited theoretical basis
renders its practical use largely an art.
PERK instead learns 
a much simpler estimator architecture
that is a weighted combination 
of nonlinear \emph{kernel} functions. 
For appropriate kernel choices,
the PERK estimator is uniquely and globally optimal
over a very diverse function space,
so PERK enjoys good expressivity
despite its simple form. 
We previously demonstrated good PERK performance
for single-compartment $\To,\Tt$ estimation 
from SS \citep{nataraj:17:dfm,nataraj:18:dfm}
and MR fingerprinting data \citep{nataraj:17:slw};
the results herein suggest
that PERK is suitable 
for higher-dimensional estimation problems as well.

\section{Methods}
\label{m,meth}

This section describes 
our myelin water imaging experiments.
\S\ref{m,meth,acq} 
applies Bayesian scan design
(\cf \S\ref{m,theory,acq})
to design a fast combination
of SPGR and DESS scans
that enables precise $\ff$ estimation.
Interestingly,
the optimized acquisition
consists only of DESS scans
(see \S\ref{m,disc} 
for discussion of this).
\S\ref{m,meth,sim}, \S\ref{m,meth,invivo}, and \S\ref{m,meth,exvivo}
describe comparisons 
respectively in simulation, \invivo, and \exvivo studies
of PERK $\ff$ estimates
from the optimized DESS acquisition
to nonnegative least-squares (NNLS) 
and regularized NNLS (RNNLS) 
conventional MWF $\mwf$ estimates
from a MESE acquisition.
\S\ref{s,est} provides PERK and conventional estimator
implementation details.
In the interest of reproducible research,
code and data will be freely available
at \url{https://github.com/gopal-nataraj/mwf}.
	
\subsection{Acquisition Design}
\label{m,meth,acq}

We assumed identical broadening distributions
across compartments
(\ie, $\Rtpf \equiv \Rtps$ and $\ompfmed \equiv \ompsmed$)
to simplify scan design optimization.
Specifically,
this simplification enabled generation 
via the \matlab Symbolic Toolbox
of cumbersome but analytical expressions
for relevant gradients and mixed gradients
of the SPGR \citep{spencer:00:mos}
and DESS \eqref{eq:model,dess-ref}-\eqref{eq:model,dess-def} 
magnitude signal models.
We used magnitude signal models
to reduce SPGR/DESS signal dependence
on off-resonance effects,
noting that Rician distributed noise 
in corresponding magnitude image data
is well-approximated as Gaussian
for sufficiently large SNR \citep{gudbjartsson:95:trd}. 
We modeled each flip angle $\flip \gets \flipnom \stx$
to deviate from a nominal prescribed value $\flipnom$ 
by spatially-varying transmit field sensitivity $\stx \approx 1$. 
We fixed $\TE$ across scans
and thereby reduced model dependencies 
to seven free object parameters per voxel:
$\ff$, $\tf{1}$, $\tf{2}$, $\ts{1}$, $\ts{2}$, $\stx$, and
$c := \mzero e^{-\Rtpf \TE} \equiv \mzero e^{-\Rtps \TE}$;
and two acquisition parameters per dataset:
$\bmp_d \gets \brac{\flipnom, \TR}\tpose, \forall d \in \set{1,\dots,D}$.
We assumed prior knowledge 
of transmit field sensitivity $\bmnu \gets \stx$ 
(that in experiments we estimated 
from separate fast acquisitions \citep{sacolick:10:bmb}-\citep{sun:14:reo})
and collected the remaining $L \gets 6$ latent parameters
as $\bmx \gets \brac{\ff, \tf{1}, \tf{2}, \ts{1}, \ts{2}, c}$.

We took fast-relaxing compartmental fraction $\ff$ 
to be a quantitative measure 
of myelin water content 
and tailored scan design problem \eqref{eq:acq,P-star}
to encourage scan combinations
that enable precise $\ff$ estimation.
Specifically,
we set weight matrix 
$\bmW \gets \diag{\inv{\expect{\bmx,\bmnu}{\ff}}, \zeros{5}}$
to penalize $\ff$ imprecision only.
Here, 
fast-fraction variance weight 
$\inv{\expect{\bmx,\bmnu}{\ff}}$ assigns interpretable meaning
to $\sqrt{\expcost\paren{\bmP}}$ 
as a unitless measure
of the expected coefficient of variation
afforded by $\bmP$ 
in asymptotically unbiased estimates of $\ff$. 

We approximated expectations 
of form $\expect{\bmx,\bmnu}{\cdot}$
by taking empirical averages
using samples of $\bmx,\bmnu$ 
drawn from a prior distribution.
We used a coordinate-wise separable prior distribution,
modeling $\ff \sim \unif{0.03, 0.21}$
to conservatively contain
state-of-the-art MESE MWF measurements in WM
\citep{zhang:15:com}
and modeling $\tf{1}$, $\tf{2}$, $\ts{1}$, and $\ts{2}$
to be Gaussian distributed
with means $400$ms, $20$ms, $1000$ms, and $80$ms
selected from literature measurements 
\citep{mackay:94:ivv, deoni:11:com}
and standard deviations
that are $20$\% of corresponding means.
Since $\bmW$ placed zero weight 
on estimating $c$, 
it sufficed to fix $c \gets 1$
and to assign noise variance
$\bmSig \gets \paren{1.49 \times 10^{-7}}\eye{10}$
based on separate measurements
in unit-normalized image data.
Lastly, 
we modeled $\stx \sim \unif{0.9,1.1}$ 
to account for $10\%$ transmit field variation
(simulations demonstrate good performance
even with $20\%$ transmit field variation).

We constrained our search space $\setP$
to reflect hardware, safety, and model-accuracy limitations
and to avoid undesirably long acquisitions.
To control RF energy deposition,
we restricted DESS flip angles 
to range between $\degrees{1}$ and $\degrees{60}$.
We further restricted SPGR flip angles
to be between $\degrees{1}$ and $\degrees{40}$
to avoid excessive model mismatch 
due to partial spoiling effects \citep{zur:91:sot}.
To comply with other fixed pulse sequence timing requirements,
we required DESS and SPGR repetition times
to be no less
than $17.5$ms and $11.8$ms respectively.
We constrained each pair
of DESS defocusing- and refocusing-echo datasets
to be assigned the same flip angle and repetition time.
Lastly, 
we imposed a somewhat ambitious total scan time constraint
$\sum_{d=1}^D \TRd \leq 108$ms
that ensured all feasible points described acquisitions
at least as fast
as the state-of-the-art SS acquisition
proposed in \citep{deoni:11:com}.
These constraints together defined a convex search space
over which we optimized $\expcost$.

We separately optimized \eqref{eq:acq,expcost}
for each of the $25$ candidate SPGR/DESS scan combinations
that are feasible under the above time constraint
and also produce at least $6$ measurements
(necessary for well-conditioned estimation).
For a candidate combination containing $D$ datasets,
we separately solved \eqref{eq:acq,P-star}
with $200D$ initializations
selected uniformly randomly 
within the feasible set.
For each combination
and each initialization,
we solved \eqref{eq:acq,P-star}
using the built-in \matlab function \texttt{fmincon}
with the \texttt{active-set} algorithm,
a maximum of $500$ iterations,
and otherwise default options.
We performed scan optimization
running \matlab R2017a 
with a pool of 12 workers
on two Xeon-X5650 2.67GHz hexa-core CPUs.

\begin{table}[!tb]
  \centering
  \begin{tabular}{r | c | c}
    \hline
    \hline
    & Optimized flip angles (deg) & Optimized repetition times (ms) \\
    \hline
    SPGR & -- 														& -- \\
    DESS & $\brac{33.0,18.3,15.1}\tpose$ 	& $\brac{17.5,30.2,60.3}\tpose$ \\
    \hline
    \hline
  \end{tabular}
  \caption{
		SPGR/DESS flip angles and repetition times
		that comprise $\est{\bmP}$,
		a scan parameter matrix designed
		under total time budget 
		$\sum_{d=1}^D \TRd \leq 108$ms
		for precise $\ff$ estimation in WM.
		For our noise variance measurements,
		this acquisition is expected
		to yield $42.5$\% coefficient of variation
		in asymptotically unbiased $\ff$ estimates
		from two-compartment signal models.
		Interestingly,
		the optimized scan combination
		consists only of DESS scans.
  }
  \label{tab:acq}
\end{table}

Table~\ref{tab:acq} summarizes
the optimized scan parameter $\est{\bmP}$ 
that locally minimizes \eqref{eq:acq,expcost}
over all combinations and all initializations.
We find that $\sqrt{\expcost(\est{\bmP})} = 0.425$, 
meaning that at a realistic noise level,
the acquisition defined by $\est{\bmP}$
is expected to yield $42.5$\% coefficient of variation
in asymptotically unbiased $\ff$ estimates
from non-exchanging two-compartment signal models.

\subsection{Simulation Studies}
\label{m,meth,sim}

We simulated data 
to arise from two non-exchanging water compartments
with different fast $\tf{2} \gets 20$ms
and slow $\ts{2} \gets 80$ms
transverse relaxation time constants
(selected based on \citep{mackay:94:ivv,deoni:11:com})
but the same bulk longitudinal relaxation time constant
$\To \gets 832$ms in WM
and $\To \gets 1331$ms in GM
(selected based on \citep{wansapura:99:nrt}).
With this two-compartment ground truth, 
MWF $\mwf$ and fast-relaxing fraction $\ff$ are equivalent,
so their estimates are statistically comparable.
(\S\ref{s,sim,3comp} describes 
a more realistic three-compartment study
where $\mwf$ and $\ff$ are no longer equivalent.)
We assigned 
$\mwf \equiv \ff \gets 0.15$ in WM
and 
$\mwf \equiv \ff \gets 0.03$ in GM
and constrained corresponding slow-compartment fractions
as $1-\ff$.
We prescribed these parameter values
to the $81$st slice
of the BrainWeb digital phantom \citep{collins:98:dac}
to produce $217 \times 181$ ground truth parameter maps.
We generated $\stx$ to model $20$\% flip angle spatial variation.
Using extended phase graphs 
to model stimulated echo signal contributions
due to non-ideal refocusing,
we simulated noiseless single-coil $32$-echo MESE image data
with nominal $\degrees{90}$ excitation 
and nominal $\degrees{180}$ refocusing flip angles;
$\TE \gets 10$ms echo interval spacing; 
$\TR \gets 600$ms repetition time;
and two cycles of gradient dephasing 
before and after each refocusing pulse.
We corrupted noiseless MESE images
with additive complex Gaussian noise
to yield noisy complex datasets
with SNR ranging 
from 17-868 in WM
and 15-697 in GM,
where SNR is defined
\begin{align}
	\snr{\bmyt,\bmepst} := \norm{\bmyt}_2/\norm{\bmepst}_2
  \label{eq:meth,snr}
\end{align}
for image data voxels $\bmyt$ and noise voxels $\bmepst$
corresponding to a region of interest (ROI)
within one image.

Using non-exchanging two-compartment models 
\eqref{eq:model,dess-ref}-\eqref{eq:model,dess-def},
we also simulated noiseless single-coil DESS image data
using the precision-optimized nominal flip angles and repetition times
presented in Table~\ref{tab:acq}
and fixed symmetric defocusing and refocusing echo times $\TE \gets 5.29$ms.
We corrupted noiseless DESS images
with additive complex Gaussian noise
to yield noisy complex datasets
with SNR ranging
from 22-222 in WM 
and 25-242 in GM,
where SNR is computed via \eqref{eq:meth,snr}.

We performed simulations and experiments
running \matlab R2013a 
on a 3.5GHz desktop computer with 32GB RAM.
We estimated $\mwf$ 
from noisy magnitude MESE images
and known $\To,\stx$ maps
by solving NNLS \eqref{eq:meth,nnls}
and RNNLS \eqref{eq:meth,rnnls} problems
as explained in \S\ref{s,est,mese}.
We estimated $\ff$ 
from noisy magnitude DESS images
and known $\stx$ maps
as detailed in \S\ref{s,est,dess}.
NNLS and RNNLS respectively took $40.3$s and $49.6$s.
PERK training and testing 
respectively took $33.8$s and $1.0$s.

\subsection{\Invivo Studies}
\label{m,meth,invivo}

We acquired all datasets 
using the TOPPE pulse sequence prototyping environment 
\citep{nielsen:18:taf}
on a GE Discovery\tmark MR750 3.0T scanner
with a 32-channel Nova Medical\regis receive head array.
In a single scan session 
involving a healthy volunteer,
we collected the optimized DESS acquisition
described in \S\ref{m,meth,acq};
a MESE acquisition for validation;
an SPGR acquisition
for separate bulk $\To$ estimation;
and a Bloch-Siegert (BS) acquisition
for separate $\stx$ estimation.
Each of these acquisitions is described next in turn.

We acquired DESS data
by prescribing the optimized nominal flip angles and repetition times
presented in Table~\ref{tab:acq}
and holding all other scan parameters fixed
across DESS scans.
For excitation,
all acquisitions in this work
used a $9.0$mm slab-selective Shinnar-Le Roux (SLR) pulse 
\citep{pauly:91:prf}
of duration~3.0ms and time-bandwidth product~6.
We interleaved RF excitations 
with 2 gradient dephasing phase cycles
over a 3mm slice thickness
to distinguish the DESS echoes.
We acquired DESS data
with a $200\times200\times8$ matrix
over a $240\times240\times24$mm$^3$ field of view (FOV).
Using a $31.25$kHz readout bandwidth,
we acquired 3D axial DESS data at minimum $\TE \gets 5.29$ms
before and after RF excitations.
To avoid slice-profile effects,
we sampled $\bmk$-space over a 3D Cartesian grid.
Including time to reach steady-state,
the optimized DESS acquisition took $3$m$15$s scan time.

We acquired MESE data
with a $90^\circ$ nominal excitation flip angle,
achieved by scaling the same SLR pulse shape as above.
A sequence of 32 identical nominally $180^\circ$ refocusing pulses
succeeded each excitation,
where the time between excitation and first refocusing pulse peaks
was fixed to the minimum possible $\frac{\TE}{2} \gets 4.6$ms
and subsequent refocusing pulse peaks were separated
by echo spacing $\TE \gets 9.2$ms.
We designed each refocusing pulse
as a $21.0$mm slab-selective SLR pulse
of duration $2.0$ms 
and time-bandwidth product~2.
We elected to use shaped refocusing pulses
instead of shorter hard pulses
to suppress unwanted signal outside the excitation slab
due to imperfect refocusing.
To suppress stimulated echo signal contributions,
we flanked each refocusing pulse
with a symmetric gradient crusher pair,
where each crusher imparted $14$ phase cycles
across the $21.0$mm refocusing slab.
Immediately following the refocusing pulse train,
we imparted $8$ gradient dephasing phase cycles 
over a $3$mm slice thickness
to suppress residual transverse magnetization.
To reduce scan time,
we used a repetition time $\TR \gets 600$ms
that is shorter than those used
in recent works (\eg, \citep{prasloski:12:rwc,zhang:15:com})
and used separate bulk $\To$ estimates
(from the SPGR scans described next)
to account for incomplete recovery.
We acquired 3D MESE data 
over the same imaging volume
and with the same resolution, readout bandwidth, and $\bmk$-space trajectory
as was used for the DESS acquisition.
We repeated the MESE scan twice
to permit averaging in postprocessing 
for increased SNR.
Including three prepended repetitions to approach steady-state,
each MESE scan took $16$m$2$s
for a total MESE acquisition time of $32$m$4$s.

We acquired SPGR data
for separate bulk $\To$ estimation,
for the sole purpose 
of aiding MESE MWF estimation.
We varied nominal flip angles 
from $5^\circ$ to $45^\circ$
with $5^\circ$ increments
and fixed all other scan parameters 
across scans.
We acquired 3D data
at minimal echo time $\TE \gets 5.1$ms
and repetition time $\TR \gets 13.1$ms
over the same imaging volume
and with the same resolution, readout bandwidth, and $\bmk$-space trajectory
as was used in the DESS acquisition.
We implemented RF spoiling
by imparting $8$ gradient dephasing phase cycles
over a $3$mm slice thickness 
immediately following each readout
and by RF phase cycling 
with a $117^\circ$ linear RF phase increment \citep{zur:91:sot}.
Including time to reach steady-state,
the SPGR acquisition took $3$m$32$s scan time.

We acquired a pair 
of BS-shifted SPGR scans \citep{sacolick:10:bmb}
for separate flip angle scaling $\stx$ estimation.
We modified the 3D SPGR sequence just described
by inserting a $\pm$4kHz off-resonant Fermi pulse
(of $9.0$ms duration
and with $0.05$G peak amplitude)
immediately following on-resonant excitation
and immediately prior to readout.
This extended the echo time to $\TE \gets 15.0$ms.
We also conservatively extended the repetition time to $\TR \gets 300$ms
to prevent excess RF heating.
We used a small $5^\circ$ nominal excitation flip angle
for reduced contrast in BS images
and thereby smoother $\stx$ estimates.
We acquired BS data
with a reduced $200\times50\times8$ matrix.
All other scan parameters were the same as for the SPGR acquisition.
Including time to reach steady-state,
the BS acquisition took $4$m$30$s scan time.

We reconstructed all raw coil images
via 3D Fourier transform
and subsequently processed only one image slice
centered within the excitation slab.
We upsampled BS coil images 
along the phase-encoding direction
to the same image size as other coil images,
using zero-padding to suppress ringing.
We jointly coil-combined all coil images
using an extension of JSENSE \citep{ying:07:jir}
for multiple datasets.
We estimated flip angle spatial variation $\stx$ maps
by normalizing and calibrating
regularized transmit field estimates \citep{sun:14:reo}
from complex coil-combined BS images.
We estimated bulk $\To$ maps
from magnitude coil-combined SPGR images and $\stx$ maps
using variable projection method \citep{golub:03:snl} 
and a one-dimensional grid search 
over $1000$ logarithmically-spaced $\To$ samples 
between $10$ms and $3000$ms.
To address bulk motion between acquisitions,
we rigidly registered coil-combined MESE and DESS images
as well as $\stx,\To$ maps
to one coil-combined MESE first-echo image.
After registration,
we averaged MESE images voxel-by-voxel
across scan repetitions
to increase effective SNR.
We estimated $\mwf$ 
from magnitude averaged MESE images and $\stx,\To$ maps 
by solving NNLS \eqref{eq:meth,nnls}
and RNNLS \eqref{eq:meth,rnnls} problems
as explained in \S\ref{s,est,mese}.
We estimated $\ff$
from magnitude DESS images and $\stx$ maps
by applying PERK
as explained in \S\ref{s,est,dess}.
NNLS and RNNLS respectively took $47.2$s and $115.4$s.
PERK training and testing respectively took $35.6$s and $0.9$s.

\subsection{\Exvivo Studies}
\label{m,meth,exvivo}

\Exvivo experiments
used a sample 
from the post-mortem brain
of an 81-year-old male
with a clinical history 
of amyotrophic lateral sclerosis (ALS).
The brain was fixed 
in 10\% neutral-buffered formalin
hours after extraction
and was sectioned 
after two weeks of fixation. 
We prepared an imaging phantom
by submerging a $\sim$1cm-thick sample 
from the prefrontal cortex 
in a viscous perfluropolyether solution
(Fomblin Y LVAC 25/6, California Vacuum Technology, Fremont, CA)
that does not produce significant MR signal 
and has been used in other studies 
(\eg, \citep{itskovich:04:qoh, symons:17:dca}).
To maintain complete immersion
and to reduce gradient-induced motion,
we anchored the sample 
at three suture sites
to our glass container
using nylon thread
(marked in Fig.~\ref{fig:exvivo}).
We waited several hours 
after phantom preparation and before scanning
to allow the phantom to equilibrate
to scan room temperature.

Similar to \invivo experiments,
we acquired DESS, MESE, SPGR, and BS data
in a single scan session.
We averaged over four MESE scan repetitions
and extended MESE repetition interval $\TR\gets1000$ms 
to further increase effective MESE SNR.
We reduced BS repetition interval $\TR\gets70.4$ms
since RF heating is of reduced concern here.
For all four acquisitions,
we collected $\bmk$-space data 
over a reduced $200\times120\times8$ grid
and reconstructed images 
onto a smaller $120\times120\times24$ mm$^3$ FOV.
To accommodate elevated apparent MWF
and shorter myelin water $\Tt$ 
after formalin fixation \citep{shatil:18:qev},
we trained PERK 
using modified $\ff,\tf{2},\ts{2}$ marginal distributions
$\dist{\ff} \gets \unif{-0.1,0.7}$,
$\dist{\tf{2}} \gets \logunif{10,30}$ms,
and
$\dist{\ts{2}} \gets \logunif{30,300}$ms.
All other data acquisition,
image reconstruction, 
and parameter estimation details
were unchanged from \invivo experiments.
Each MESE repetition took $16$m$3$s
for a total MESE acquisition time of $64$m$12$s.
DESS, SPGR, BS acquisitions 
respectively took $1$m$57$s, $2$m$7$s, and $2$m$32$s. 
NNLS and RNNLS respectively took $28.0$s and $34.8$s.
PERK training and testing respectively took $35.8$s and $0.2$s.

\section{Results}
\label{m,result}

This section demonstrates myelin water imaging
using our precision-optimized DESS acquisition
(\cf \S\ref{m,theory,acq}, \ref{m,meth,acq})
and fast PERK estimation 
(\cf \S\ref{m,theory,perk}, \ref{s,est,dess}).
\S\ref{m,result,sim}, \S\ref{m,result,invivo}, and \S\ref{m,result,exvivo}
compare DESS PERK $\ff$ estimates 
to conventional MESE NNLS/RNNLS $\mwf$ estimates
from the simulation, \invivo, and \exvivo studies
respectively described 
in \S\ref{m,meth,sim}, \S\ref{m,meth,invivo}, and \S\ref{m,meth,exvivo}.

\subsection{Simulation Studies}
\label{m,result,sim}

\begin{figure}[!t]
	\centering
	\subfigure{%
		\includegraphics [width=\textwidth] {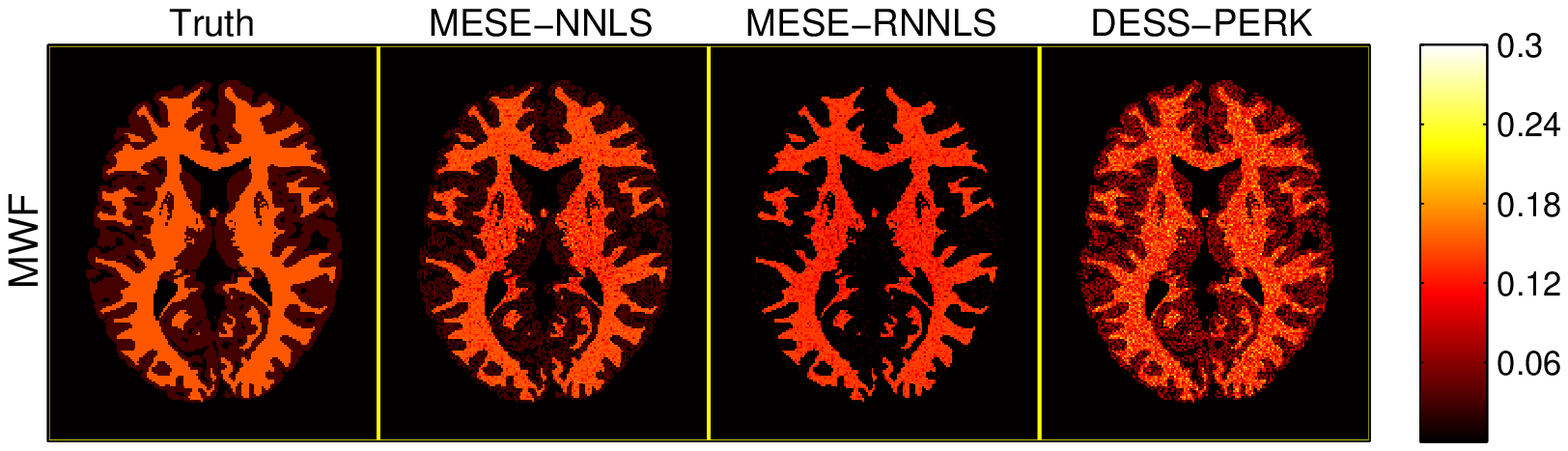}
		\label{fig:2comp,im}
	}
	\hspace{0cm}
	\subfigure{%
		\includegraphics [width=\textwidth,trim=0 0 0 25,clip] {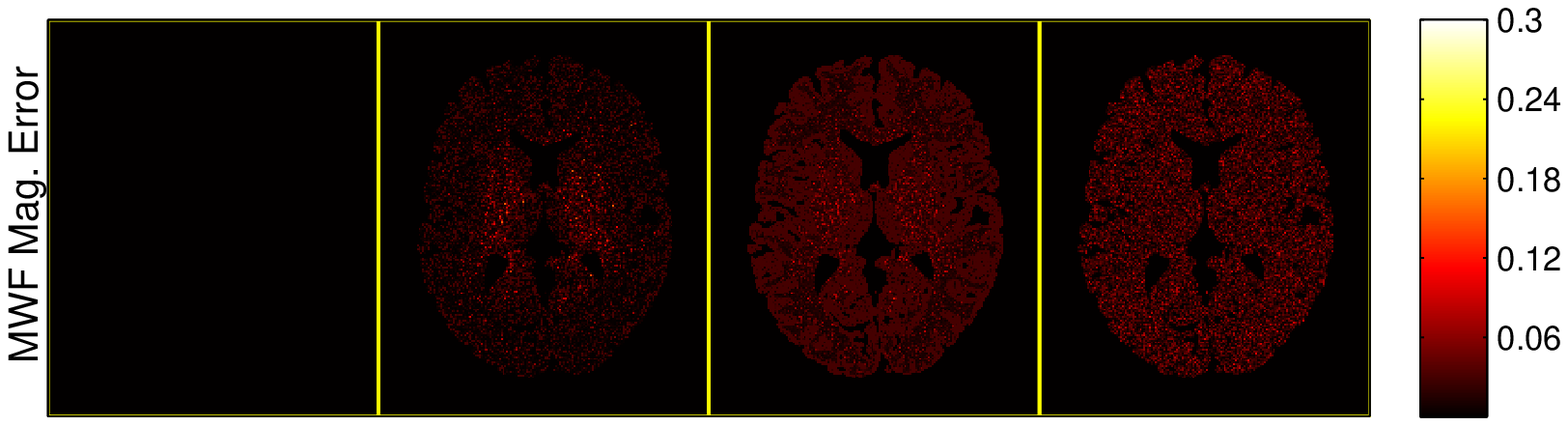}
		\label{fig:2comp,err}
	}
	\caption{%
		MESE NNLS/RNNLS $\mwf$ estimates 
		and DESS PERK $\ff$ estimates 
		alongside corresponding magnitude error images,
		in a two-compartment simulation.
		Voxels not assigned WM- or GM-like compartmental fractions
		are masked out in post-processing for display.
		Table~\ref{tab:2comp} presents corresponding sample statistics.
	}
	\label{fig:2comp}
\end{figure}

Fig.~\ref{fig:2comp} compares MESE NNLS/RNNLS $\mwf$ 
and DESS PERK $\ff$ estimates
alongside magnitude difference images
with respect to the ground truth $\mwf \equiv \ff$ map.
Unlike the DESS $\ff$ estimate,
both $\mwf$ estimates visibly exhibit systematic error
due to flip angle spatial variation
despite perfect knowledge of $\stx$;
this apparent MESE sensitivity 
to transmit field variation increases
in the presence 
of realistic model mismatch
(see \S\ref{s,sim,3comp}).

\begin{table}[!t]
	\centering
	\begin{tabular}{r | r r}
		\hline
		\hline
													& WM 															& GM \\
		\hline
		True $\mwf\equiv\ff$	& $0.15$ 													& $0.03$ \\
		\hline
		MESE-NNLS $\mwfest$ 	&	\mnstd{0.1375}{0.0187} (0.0225) & \mnstd{0.0203}{0.01296} (0.0162) \\
		MESE-RNNLS $\mwfest$ 	& \mnstd{0.1285}{0.0146} (0.0260) & \mnstd{0.00207}{0.00524} (0.02841) \\
		DESS-PERK $\ffest$ 		& \mnstd{0.1352}{0.0267} (0.0305) & \mnstd{0.0436}{0.0267} (0.0299) \\
		\hline
		\hline
	\end{tabular}
	\caption{%
		Sample means $\pm$ sample standard deviations (RMSEs)
		of MESE NNLS/RNNLS $\mwf$ estimates
		and DESS PERK $\ff$ estimates
		in a two-compartment simulation.
		Sample statistics are computed 
		over $7810$ WM-like and $9162$ GM-like voxels.
		Each sample statistic is rounded off 
		to the highest place value
		of its (unreported) standard error,
		computed via formulas in \citep{ahn:03:seo}.
		Fig.~\ref{fig:2comp} presents corresponding images.
	}
	\label{tab:2comp}
\end{table}

Table~\ref{tab:2comp} compares sample statistics
of MESE NNLS/RNNLS $\mwf$ estimates
and DESS PERK $\ff$ estimates,
computed over $7810$ WM-like and $9162$ GM-like voxels.
Except for the MESE-RNNLS GM $\mwf$ estimate,
all estimates agree with true values
to within one standard deviation.
The MESE-NNLS WM and GM $\ff$ estimates 
achieve the least root mean-squared errors (RMSEs) overall.
The RNNLS $\mwf$ estimate is more precise but less accurate
than the NNLS $\mwf$ estimate 
due to regularization.
To better assess 
whether PERK is suitable 
for DESS $\ff$ estimation,
\S\ref{s,sim,2comp} extends this simulation
by adding conventional grid search $\ff$ estimation
(that is practical only in simulation).

\subsection{\Invivo Studies}
\label{m,result,invivo}

\begin{figure}[!t]
	\centering
	\includegraphics [width=\textwidth] {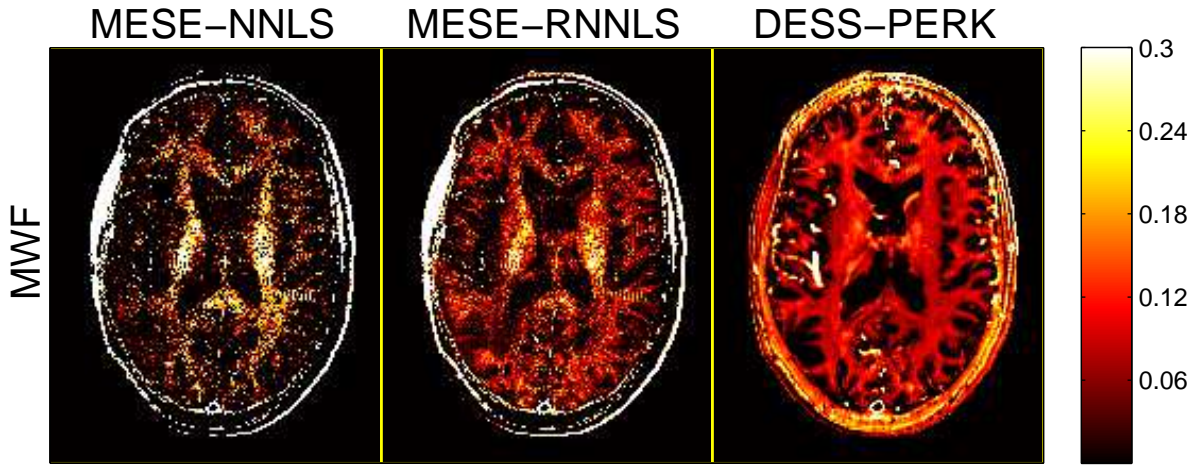}
	\caption{%
		Representative NNLS and RNNLS $\mwf$ estimates 
		from a MESE acquisition 
		alongside a PERK $\ff$ estimate 
		from a precision-optimized DESS acquisition,
		in the brain of a healthy volunteer.
		Using similar signal reception imaging parameters,
		MESE $\mwf$ estimates 
		took $40$m$6$s total scan time
		(averaging over two MESE scan repetitions)
		while DESS $\ff$ estimates
		took $7$m$45$s total scan time. 
		PERK $\ff$ estimates exhibit less WM variation
		and more clearly delineate cortical WM/GM boundaries
		than MESE $\mwf$ estimates.
		Table~\ref{tab:invivo} 
		presents corresponding sample statistics
		computed over manually selected WM and GM ROIs.
	}%
	\label{fig:invivo}
\end{figure}

Fig.~\ref{fig:invivo} compares 
NNLS and RNNLS $\mwf$ estimates from MESE scans
as well as PERK $\ff$ estimates from optimized DESS scans.
PERK $\ff$ estimates exhibit less WM variation
and more clearly delineate cortical WM/GM boundaries
than MESE $\mwf$ estimates.
RNNLS $\mwf$ estimates are visibly lower than NNLS $\mwf$ estimates
due to regularization
but exhibit reduced WM variation,
somewhat improving visualization of WM tracts. 
RNNLS $\mwf$ and PERK $\ff$ estimates 
appear visually similar in lateral WM regions,
but both NNLS and RNNLS $\mwf$ estimates are elevated in medial regions.
Elevated MESE $\mwfest$ estimates 
in internal capsules (IC)
have been attributed 
to overlapping myelin water and cellular water $\Tt$ peaks
in MESE $\Tt$ spectrum estimates \citep{zhang:15:com}.
We additionally observe
that MESE $\mwf$ estimates exhibit similar spatial variation
here versus in simulations 
(\cf Figs.~\ref{fig:2comp-ext}-\ref{fig:3comp-ext})
suggesting that some spatial variation
in MESE $\mwf$ estimates
may be attributable in part to flip angle variation,
despite compensation for transmit field inhomogeneity.

\begin{table}[!t]
	\centering
	\small
	\begin{minipage}{0.19\textwidth}
		\includegraphics [width=\textwidth] {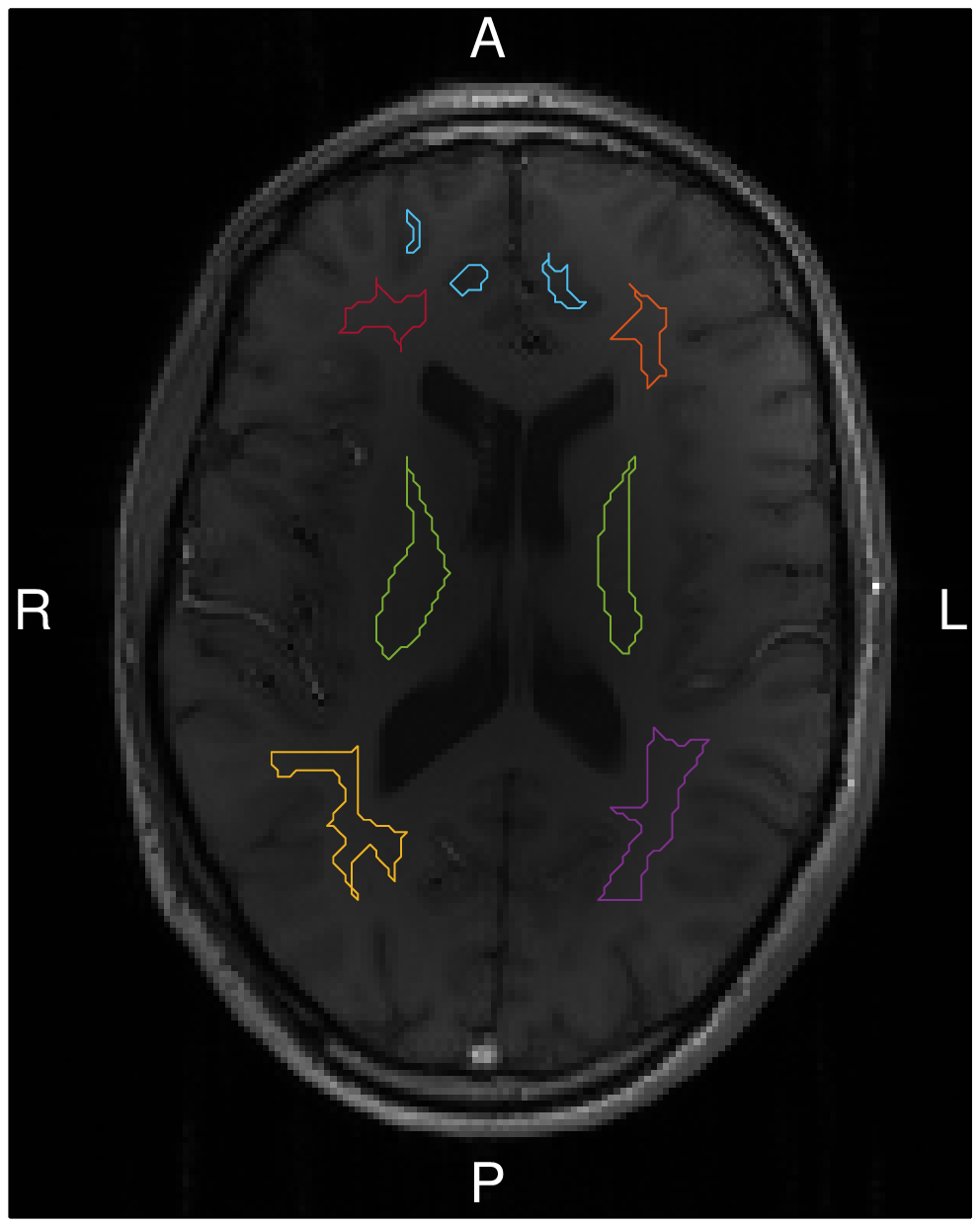}
	\end{minipage}
	\begin{minipage}{0.78\textwidth}
		\begin{tabu} {c | r r r}
			\hline
			\hline
			ROI 		& MESE-NNLS $\mwfest$ 	& MESE-RNNLS $\mwfest$	& DESS-PERK $\ffest$ 	\\
			\hline
			\RA WM 	& \mnstd{0.081}{0.091}  & \mnstd{0.074}{0.054} 	& \mnstd{0.117}{0.019}	\\
   		\LA WM 	& \mnstd{0.068}{0.086}  & \mnstd{0.054}{0.041}  &	\mnstd{0.100}{0.0119}	\\
   		\RP WM 	& \mnstd{0.049}{0.075}  & \mnstd{0.043}{0.041} 	&	\mnstd{0.093}{0.019}	\\
   		\LP WM 	& \mnstd{0.118}{0.095}  & \mnstd{0.075}{0.050}  &	\mnstd{0.0870}{0.0114}\\
  		\IC WM 	& \mnstd{0.208}{0.133}  & \mnstd{0.177}{0.083}  &	\mnstd{0.111}{0.0241}	\\
 			\AC GM 	& \mnstd{0.005}{0.020}  & \mnstd{0.009}{0.017} 	&	\mnstd{0.019}{0.045}	\\
			\hline
			\hline
		\end{tabu}
	\end{minipage}
	\caption{%
		\emph{Left}:
		WM/GM ROIs,
		overlaid on a representative MESE first-echo image.
		Separate lateral WM ROIs are distinguished
		by anterior-right (\RA),
		anterior-left (\LA),
		posterior-right (\RP),
		and posterior-left (\LP) directions
		and are respectively comprised
		of $90$, $79$, $182$, and $201$ voxels.
		Two internal capsule (\IC) polygons
		are pooled into a single medial WM ROI
		comprised of $347$ voxels.
		Three small anterior cortical (\AC) GM polygons
		are pooled into a single GM ROI
		comprised of $78$ voxels.
		\emph{Right}:
		Sample means $\pm$ sample standard deviations
		of NNLS/RNNLS $\mwf$ estimates 
		from a MESE acquisition
		as well as PERK $\ff$ estimates 
		from an optimized DESS acquisition,
		computed over WM/GM ROIs.
		Each sample statistic is rounded off
		to the highest place value
		of its (unreported) standard error,
		computed via formulas in \citep{ahn:03:seo}.
		Fig.~\ref{fig:invivo} presents corresponding images.
	}%
	\label{tab:invivo}
\end{table}

Table~\ref{tab:invivo} summarizes sample statistics
of NNLS/RNNLS $\mwf$ estimates from MESE scans
and PERK $\ff$ estimates from optimized DESS scans,
separately computed over four lateral WM ROIs
containing $90$, $79$, $182$, and $201$ voxels;
one pooled medial IC WM ROI containing $347$ voxels;
and one pooled anterior cortical (AC) GM ROI containing $78$ voxels.
PERK $\ff$ estimates exhibit the lowest variation within WM ROIs
and the most similar sample means across WM ROIs.
NNLS and RNNLS $\mwf$ sample means are significantly higher 
in the IC WM ROI than in lateral WM ROIs,
possibly due to overlap in NNLS $\Tt$ spectrum peaks
and/or to flip angle spatial variation
(as described in the previous paragraph).
It is challenging to assess quantitative comparability
between DESS $\ff$ and MESE $\mwf$ ROI sample means
due to high within-ROI variation in MESE estimates,
though PERK WM $\ff$ sample means agree reasonably
with several other methods
(\eg, see \citep{alonsoortiz:15:mbm} for a review).
Neither the NNLS/RNNLS $\mwfest$ nor PERK $\ffest$ estimators
measured significant myelin water content in AC GM.

\subsection{\Exvivo Studies}
\label{m,result,exvivo}

\begin{figure}[!t]
	\centering
	\includegraphics [width=\textwidth] {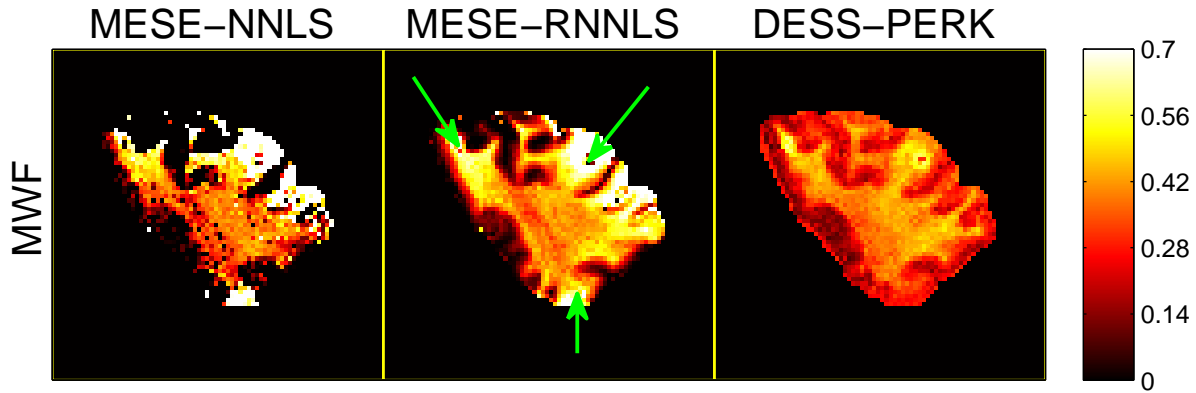}
	\caption{%
		NNLS and RNNLS $\mwf$ estimates 
		from a MESE acquisition
		alongside a PERK $\ff$ estimate 
		in a formalin-fixed sample
		from the prefrontal cortex of an ALS patient 
		about three weeks after death.
		Using similar signal reception imaging parameters,
		MWF $\mwf$ estimates 
		took $68$m$51$s total scan time
		(averaging over four MESE scan repetitions)
		while DESS $\ff$ estimates
		took $4$m$29$s total scan time.
		Green arrows indicate three suture sites 
		used to anchor and stabilize the sample. 
		Warping of the $\sim$1cm-thick sample
		caused partial-volume effects 
		in the anterior-left and posterior regions
		(clearly apparent
		in the anatomical MESE image 
		within Table~\ref{tab:exvivo}).
		Away from these regions, 
		MESE $\mwf$ and DESS $\ff$ estimates 
		exhibit similar spatial variation
		and are in reasonable agreement
		in WM and near WM/GM boundaries.
		Table~\ref{tab:exvivo} 
		presents corresponding sample statistics
		computed over manually selected WM and GM ROIs.
	}%
	\label{fig:exvivo}
\end{figure}

Fig.~\ref{fig:exvivo} compares 
NNLS and RNNLS $\mwf$ estimates from MESE scans
as well as 
PERK $\ff$ estimates from optimized DESS scans.
All estimates are higher 
than corresponding \invivo estimates,
likely due to formalin fixation \citep{shatil:18:qev}.
Green arrows mark three suture sites.
Warping of the $\sim$1cm-thick sample
caused partial-volume effects 
in anterior-left and posterior regions
of this reconstructed slice
(see shading in the MESE image 
within Table~\ref{tab:exvivo};
adjacent slices contained larger affected regions).
Away from suture sites 
and partial-volume affected regions,
MESE $\mwf$ and DESS $\ff$ estimates
exhibit similar spatial variation
and are in reasonable agreement
in WM and near WM/GM boundaries.
In GM,
DESS $\ff$ estimates appear higher
than MESE $\mwf$ estimates,
likely because PERK is here being trained
with a sampling distribution support 
much broader than the prior distribution 
over which DESS was designed 
to enable precise $\ff$ estimation.
In particular, 
\exvivo training marginal 
$\dist{\ff} \gets \paren{-0.1,0.7}$ 
is much broader 
than acquisition design prior $\ff \sim \unif{0.03,0.21}$. 
Narrowing $\dist{\ff}$ to the \invivo range 
improves agreement across $\ff$ and $\mwf$ estimators in \exvivo GM,
but degrades PERK performance in \exvivo WM
since PERK is then extrapolating
well beyond its training distribution.

\begin{table}[!t]
	\centering
	\small
	\begin{minipage}{0.19\textwidth}
		\includegraphics [width=\textwidth] {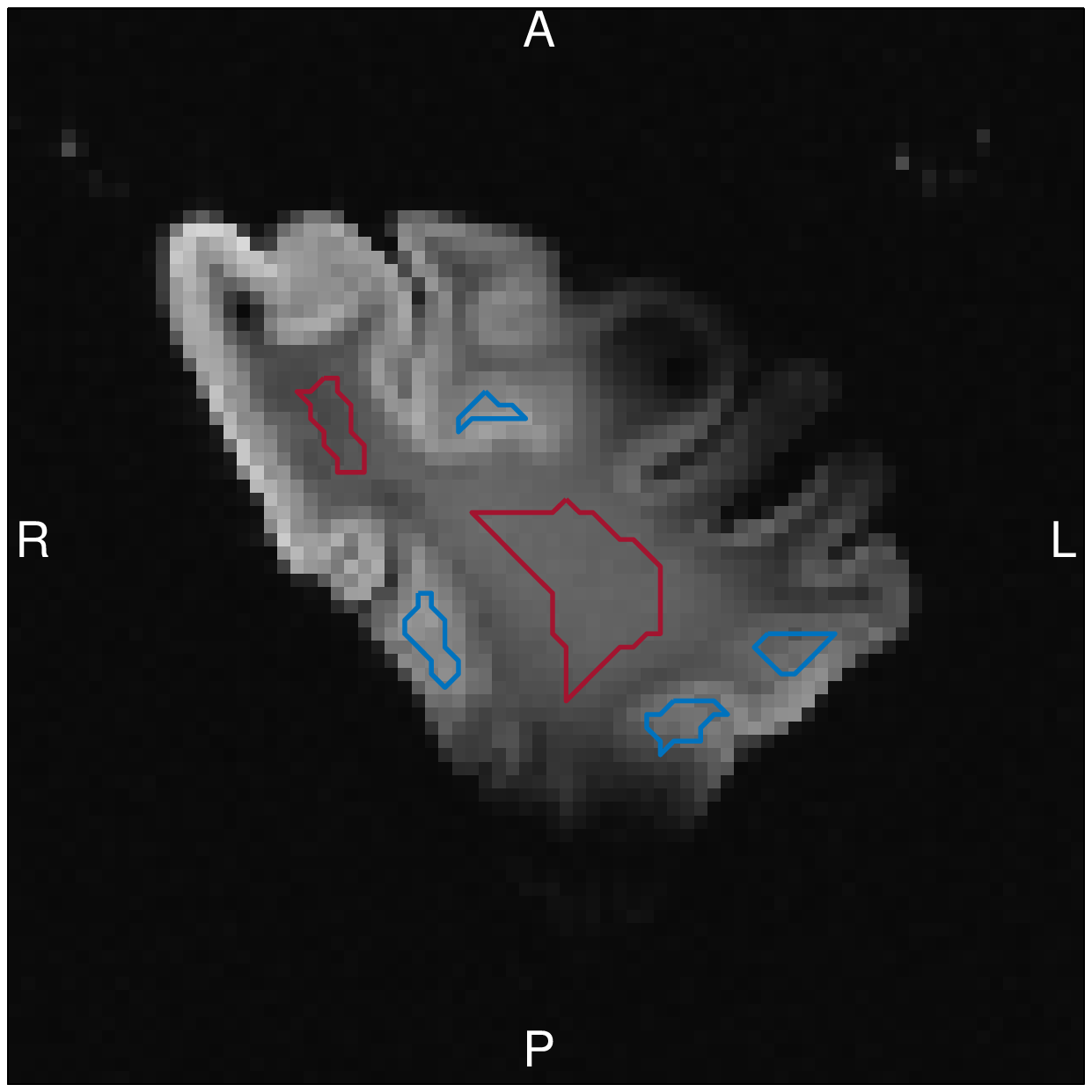}
	\end{minipage}
	\begin{minipage}{0.78\textwidth}
		\begin{tabu} {c | r r r}
			\hline
			\hline
			ROI 		& MESE-NNLS $\mwfest$ 	& MESE-RNNLS $\mwfest$	& DESS-PERK $\ffest$ 	\\
			\hline
			\WM			&	\mnstd{0.366}{0.113}	& \mnstd{0.426}{0.061}	& \mnstd{0.420}{0.029} \\
			\GM			& \mnstd{0.08}{0.102} 	& \mnstd{0.20}{0.14} 		& \mnstd{0.225}{0.075} \\
			\hline
			\hline
		\end{tabu}
	\end{minipage}
	\caption{%
		\emph{Left}:
		\WM/\GM ROIs,
		overlaid on a representative
		MESE twelfth-echo image.
		The \WM ROI consists 
		of two manually-selected polygons
		pooled into a single ROI
		consisting of $143$ voxels.
		The \GM ROI consists
		of four manually-selected polygons
		pooled into a single ROI
		consisting of $73$ voxels.
		\emph{Right}:
		Sample means $\pm$ sample standard deviations
		of NNLS/RNNLS $\mwf$ estimates 
		from a MESE acquisition
		as well as PERK $\ff$ estimates 
		from an optimized DESS acquisition,
		computed over WM/GM ROIs.
		Each sample statistic is rounded off
		to the highest place value
		of its (unreported) standard error,
		computed via formulas in \citep{ahn:03:seo}.
		Fig.~\ref{fig:exvivo} presents corresponding images.
		Within the WM ROI,
		MESE $\mwf$ and DESS $\ff$ estimates 
		are quantitatively comparable.
	}%
	\label{tab:exvivo}
\end{table}

Table~\ref{tab:exvivo} summarizes sample statistics
of NNLS/RNNLS $\mwf$ estimates from MESE scans
and PERK $\ff$ estimates from optimized DESS scans,
computed over manually-selected WM and GM ROIs 
respectively containing $143$ and $73$ voxels.
Within both ROIs,
PERK $\ff$ estimates exhibit the lowest variation.
Within the WM ROI,
MESE $\mwf$ and DESS $\ff$ estimates 
are quantitatively comparable.
Though RNNLS $\mwf$ and PERK $\ff$ sample means are similar in GM,
it is challenging to quantitatively compare 
$\mwf$ and $\ff$ estimates in GM
due to high within-ROI variation in MESE estimates.

\section{Discussion}
\label{m,disc}

Simulations and experiments demonstrate the feasibility 
of myelin water content quantification
using a precision-optimized DESS MR acquisition
and fast machine learning with kernels (PERK). 
Simulations show 
that DESS PERK $\ff$ estimators
and conventional MESE $\mwf$ estimators
achieve comparable RMSE 
in WM- and GM-like voxels.
\Invivo and \exvivo experiments demonstrate
that MESE $\mwf$ and DESS $\ff$ estimates
are quantitatively comparable measures
of WM myelin water content.
To our knowledge,
these experiments are the first 
to demonstrate myelin water images 
from a SS acquisition
that are quantitatively similar
to conventional MESE MWF images.

Despite freedom to design arbitrary combinations
of SPGR and DESS scans,
the optimized acquisition used here,
as well as several other unreported acquisitions 
designed under different total time constraints,
ended up consisting either entirely or mostly of DESS scans.
Since the two-compartment SPGR signal models 
used in acquisition design
depend on $\tf{1},\ts{1}$ 
but not $\tf{2},\ts{2}$,
DESS-dominated scan designs suggest 
that multi-compartmental $\Tt$ effects
give rise to $\ff$ sensitivity in SS sequences
more so than multi-compartmental $\To$ effects.
Perhaps surprisingly,
reported and unreported precision-optimized acquisitions 
also exhibit substantial $\TR$ diversity across scans,
even at the expense
of fewer scans than possible 
under time constraints.
In further unreported studies,
we investigated this phenomenon 
by repeating scan optimization
while implicitly constraining repetition times to be minimal.
We consistently observed substantial ($\sim$10-20\%) degradation
in expected $\ff$ coefficient of variation,
suggesting that $\TR$ diversity 
(in addition to flip angle diversity)
is important for designing acquisitions
that enable precise $\ff$ estimation.

Our experiments used non-exchanging SPGR and DESS signal models
so as to work with closed-form signal models and gradients
during acquisition design
and to keep consistent 
with standard MESE model assumptions.
There is growing evidence however
that some significant physical exchange
across the myelin bilayers
likely persists in cerebral WM
(\eg, see \citep{does::ibt} for a recent review).
A thorough investigation 
of the sensitivity 
of DESS $\ff$ or MESE $\mwf$ estimates
to realistic physical exchange rates
is a topic for further research.

Even with high SNR,
differences in model assumptions, cost functions, and estimation algorithms
may limit the quantitative comparability 
of DESS $\ff$ and MESE $\mwf$ imaging
as implemented here.
For more similar model assumptions,
one could attempt to estimate 
from a suitably optimized DESS acquisition
a $\Tt$ (or joint $\paren{\To,\Tt}$) distribution
using two-compartment, three-compartment, or higher-compartment models
and correspondingly estimate from MESE data
a more coarsely sampled $\Tt$ (or joint $\paren{\To,\Tt}$) distribution.
If $\stx,\To$ maps are known
and non-exchanging additive models are employed,
one could estimate $\Tt$ distributions
from both MESE and DESS data 
using NNLS or RNNLS objective functions.
With milder model assumptions 
that cause signal models 
to be nonlinear in unknowns,
one could instead estimate distributions using PERK.
This work focused on demonstrating the feasibility
of myelin water content quantification
using a simple two-compartment model 
of a fast DESS acquisition;
estimating more unknowns 
from more complicated models
will necessitate more scans
but could be an interesting area
for further research.

\section{Conclusion}
\label{m,conc}

This paper has introduced a new method
for precise myelin water content quantification 
based on a fast SS acquisition
and PERK \citep{nataraj:18:dfm},
a fast, scalable machine learning algorithm 
for MRI parameter estimation.
The acquisition consists
of three DESS scans
whose flip angles and repetition times
were optimized under a competitive time constraint
to enable precise estimation
of the faster-relaxing signal fraction $\ff$ 
in a two-compartment DESS signal model.
Simulations demonstrated that DESS PERK $\ff$ estimators
and conventional MESE $\mwf$ estimators
achieve comparable RMSE in WM- and GM-like voxels.
\Invivo and \exvivo experiments demonstrated
that MESE $\mwf$ and DESS PERK $\ff$ estimates
are quantitatively comparable measures
of WM myelin water content.
To our knowledge,
these experiments are the first 
to demonstrate myelin water images
from a SS acquisition
that are quantitatively similar
to conventional MESE MWF images.

\section*{Acknowledgments}
\label{b,ack}

We thank Roger Albin
for initial discussions 
about neurodegenerative applications,
Clay Scott 
for several discussions about kernel learning,
Marina Epelman 
for a discussion 
about global optimization for scan design,
and Scott Swanson
for several discussions about myelin.
We also thank the University of Michigan Brain Bank
for providing the brain tissue
used in \exvivo studies
and Tyler Cork for suggesting Fomblin
to prepare our \exvivo phantom.

\section*{Appendix}
\label{b,apx}

\subsection*{Gradient of Bayesian Scan Design Cost}
\label{b,apx,acq}

Acquisition design cost $\expcost$ 
is non-convex but typically differentiable 
in acquisition parameter matrix $\bmP$.
Here we construct the gradient matrix
$\grada{\bmP}{\expcost\paren{\bmP}} \in \reals{A \times D}$
and provide sufficient conditions
for when this gradient matrix exists.
Our strategy involves 
first constructing $\grada{\bmP}{\costa{\bmx,\bmnu,\bmP}}$
element-wise 
for fixed $\bmx,\bmnu$
and then relating
$\grada{\bmP}{\expcost\paren{\bmP}}$
to $\grada{\bmP}{\costa{\bmx,\bmnu,\bmP}}$.
Let $\dela{p_{a,d}}$ be the $(a,d)$th element
of matrix operator $\grada{\bmP}$.
By standard matrix derivative identities,
we have
\begin{align}
	\dela{p_{a,d}} \costa{\bmx,\bmnu,\bmP}
		&\equiv 
		\dela{p_{a,d}} \trace{\bmW \bmF^{-1}\paren{\bmx,\bmnu,\bmP} \bmW}
		\nonumber \\
		&= -\trace{\bmW \bmF^{-1}\paren{\bmx,\bmnu,\bmP} 
		\dela{p_{a,d}}\paren{\bmFa{\bmx,\bmnu,\bmP}} 
		\bmF^{-1}\paren{\bmx,\bmnu,\bmP} \bmW}.
		\label{eq:acq,cost-der}
\end{align}
If elements of measurement vector $\bmy$ 
are assumed to be independent
as is typical,
$\bmSig$ takes the form 
$\bmSig \gets \diag{\brac{\sigma_1^2,\dots,\sigma_D^2}\tpose}$
and
\begin{align}
	\dela{p_{a,d}}\paren{\bmFa{\bmx,\bmnu,\bmP}} 
		&= 
		\dela{p_{a,d}} \sum_{d'=1}^D \frac{1}{\sigma_{d'}^2}
		\paren{\grada{\bmx}{s_{d'}\paren{\bmx,\bmnu,\bmp_{d'}}}}\ctpose
		\grada{\bmx}{s_{d'}\paren{\bmx,\bmnu,\bmp_{d'}}}
		\nonumber \\
		&= 
		\frac{1}{\sigma_d^2} \dela{p_{a,d}} 
		\paren{\paren{\grada{\bmx}{s_d\paren{\bmx,\bmnu,\bmp_d}}}\ctpose
		\grada{\bmx}{s_d\paren{\bmx,\bmnu,\bmp_d}}},
		\label{eq:acq,fisher-der}
\end{align}
where $\diag{\cdot}$ assigns its argument
to the diagonal entries
of an otherwise zero matrix;
and $s_d$ and $\bmp_d$ 
respectively denote $d$th entry of $\bms$
and the $d$th column of $\bmP$.
Substituting \eqref{eq:acq,fisher} and \eqref{eq:acq,fisher-der}
into \eqref{eq:acq,cost-der} 
gives expressions 
in terms of signal model derivatives
for each element 
of $\grada{\bmP}{\costa{\bmx,\bmnu,\bmP}}$.
These expressions are well-defined 
if $\bmF$ is invertible 
and if mixed partial derivatives
$\grada{\bmp_1}{\paren{\grada{\bmx}{s_1}}\tpose},
\dots,
\grada{\bmp_D}{\paren{\grada{\bmx}{s_D}}\tpose}$
exist and are continuous in $\bmx,\bmP$,
where $\paren{\cdot}\tpose$ denotes transpose.
Further assuming 
that $\grada{\bmP}{\costa{\bmx,\bmnu,\bmP}}$ remains bounded 
for all $\bmx,\bmnu$, 
\begin{align}
	\grada{\bmP}{\expcost\paren{\bmP}} 
		&\equiv 
		\grada{\bmP}{\expect{\bmx,\bmnu}{\costa{\bmx,\bmnu,\bmP}}}
		\nonumber \\
		&= 
		\expect{\bmx,\bmnu}{\grada{\bmP}{\costa{\bmx,\bmnu,\bmP}}},
		\label{eq:acq,expcost-grad}
\end{align} 
which provides an expression
for the gradient of the expected cost,
as desired.

\subsection*{Brief Review of PERK}
\label{b,apx,est}

PERK learns a nonlinear estimator
from simulated labeled training points.
PERK first samples 
a prior joint distribution on $\bmx,\bmnu$ 
and evaluates signal model \eqref{eq:acq,model} $N$ times
(with previously optimized 
and now fixed acquisition parameter $\bmPstar$)
to generate sets of parameter and noise realizations
$\set{\bmx_1,\bmnu_1,\bmeps_1},
\dots,
\set{\bmx_N,\bmnu_N,\bmeps_N}$
and corresponding measurements
$\set{\bmy_1,\dots,\bmy_N}$.
PERK then seeks to learn
from these samples
a suitable regression function 
$\esta{\bmx}{\cdot} : \reals{Q} \mapsto \reals{L}$
that maps each regressor 
$\bmq_n := \brac{\abs{\bmy_n}\tpose,\bmnu_n\tpose}\tpose$
to an estimate $\esta{\bmx}{\bmq_n}$ 
that is ``close'' to corresponding regressand $\bmx_n$,
where $Q := D+K$ 
and $n \in \set{1,\dots,N}$.
This supervised learning problem
is subject to an inherent tradeoff
between training complexity
and estimator accuracy.
At one extreme, 
restricting the estimator
to the affine form 
$\esta{\bmx}{\cdot} \gets \est{\bma}\tpose\paren{\cdot}+\est{\bmb}$
(\ie, affine regression)
typically corresponds 
to well-posed training problem,
but an affine estimator is unlikely to be useful
when the signal model is nonlinear in $\bmx$.
At the other extreme,
attempting to learn an overly flexible estimator may fail
because many candidate regression functions
fit any finite $N$ training points 
with zero training error.
PERK balances between these extremes 
by learning an estimator $\est{\bmx}$
of form $\esta{\bmx}{\cdot} \gets
	\sum_{n=1}^N \est{\bma}\, k\paren{\cdot,\bmq_n} + \est{\bmb}$,
where $k : \reals{2Q} \mapsto \real$
is a (typically nonlinear) 
\emph{reproducing kernel} function \citep{aronszajn:50:tor}.
Specifically,
the PERK estimator reads
\begin{align}
	\esta{\bmx}{\cdot} \gets \bmX \paren{%
		\frac{1}{N}\ones{N} 
		+ \bmM\inv{\bmM\bmK\bmM+N\rho\eye{N}}\bmka{\cdot}
	},
	\label{eq:perk,est}
\end{align}
where $\bmX := \brac{\bmx_1,\dots,\bmx_N}$
collects the regressands;
$\ones{N} \in \reals{N}$ 
denotes a vector of ones;
$\bmM := \eye{N}-\frac{1}{N}\ones{N}\ones{N}\tpose$
denotes a de-meaning operator;
$\eye{N} \in \reals{N \times N}$ 
denotes an identity matrix;
Gram matrix $\bmK \in \reals{N \times N}$ 
consists of entries $k\paren{\bmq_n,\bmq_{n'}}$
for $n,n' \in \set{1,\dots,N}$;
$\rho>0$ is a regularization parameter; 
and 
$\bmk\paren{\cdot} := 
	\brac{k\paren{\cdot,\bmq_1},\dots,k\paren{\cdot,\bmq_N}}\tpose
	- \frac{1}{N} \bmK \ones{N}
	: \reals{Q} \mapsto \reals{N}$
is a (typically nonlinear) kernel embedding operator.

PERK estimator \eqref{eq:perk,est}
is the uniquely and globally optimal regression function
within a certain function space
whose richness is determined 
by the choice of kernel,
and for good PERK accuracy it is desirable
for this function space to be sufficiently diverse.
As in \citep{nataraj:18:dfm}, 
we use Gaussian kernel 
\begin{align}
	k(\bmq,\bmq') \gets 
		\expa{-\frac{1}{2}\norm{\bmq-\bmq'}^2_{\bmL^{-2}}},
	\label{eq:perk,kernel}
\end{align}
where $\bmL \in \reals{Q \times Q}$
denotes a positive definite bandwidth matrix
(that can be selected 
in a data-driven manner \citep{nataraj:18:dfm})
and $\norm{\cdot}_{\bmG} \equiv \norm{\bmG^{1/2}\paren{\cdot}}$ 
denotes a weighted $\ell^2$ norm 
for positive semidefinite $\bmG$. 
For this kernel choice,
PERK can approximate $\Ltwo$ functions
to arbitrary accuracy
for $N$ sufficiently large \citep{steinwart:08:svm}.

More challenging applications
typically require larger numbers of training samples $N$,
which complicates direct use
of PERK estimator \eqref{eq:perk,est}
due its dependence 
on dense $N \times N$ Gram matrix $\bmK$.
Fortunately, 
Gaussian kernel \eqref{eq:perk,kernel} 
admits an approximation 
$k\paren{\bmq,\bmq'} \approx \bmza{\bmq}\tpose\bmza{\bmq'} \,
	\forall \bmq,\bmq'$
\citep{rahimi:07:rff}
that enables constructing 
$\bmZ := \brac{\bmza{\bmq_1},\dots,\bmza{\bmq_N}} 
	\in \reals{Z \times N}$
such that $\bmZ\tpose\bmZ \approx \bmK$ 
for $Z \ll N$,
where $\bmz : \reals{Q} \mapsto \reals{Z}$ 
denotes an approximate nonlinear \emph{feature map}
that admits very fast implementation \citep{nataraj:18:dfm}.
Substituting low-rank approximation $\bmZ\tpose\bmZ$
in place of $\bmK$ 
in \eqref{eq:perk,est}
and applying the matrix inversion lemma \citep{woodbury:50:imm} 
yields approximate PERK estimator
\begin{align}
	\esta{\bmx}{\cdot} \gets
		\bmmx + \bmCxz \inv{\bmCzz + \rho\eye{Z}}
		\paren{\bmza{\cdot} - \bmmz},
	\label{eq:perk,approx}
\end{align}
where $\bmmx := \frac{1}{N}\bmX\ones{N}$ 
and $\bmmz := \frac{1}{N}\bmZ\ones{N}$ 
are sample means;
and $\bmCxz := \frac{1}{N}\bmX\bmM\bmZ\tpose$ 
and $\bmCzz := \frac{1}{N}\bmZ\bmM\bmZ\tpose$ 
are sample covariances. 
Estimator \eqref{eq:perk,approx} elucidates 
that Gaussian PERK is approximately equivalent 
to first nonlinearly transforming regressors 
$\bmq_1,\dots,\bmq_N$
into features 
$\bmza{\bmq_1},\dots,\bmza{\bmq_N}$
and then performing regularized affine regression
with these (typically higher-dimensional) features;
this approximation approaches equality
asymptotically in $Z$.

\newpage
\bibliographystyle{unsrt}
\bibliography{./bib/master}

\pagebreak
\begin{center}
	\huge{%
		Supporting Information for \\
		Fast, Precise Myelin Water Quantification \\
		using DESS MRI and Kernel Learning
	}
	\vspace{0.5cm}	
	
	\large{%
		Gopal~Nataraj$^\star$, %
		Jon-Fredrik~Nielsen$^\dagger$, %
		Mingjie~Gao$^\star$, %
		and %
		Jeffrey~A.~Fessler$^\star$
	}
	\vspace{0.2cm}
	
	\date{%
		$^\star$Dept. of Electrical Engineering and Computer Science, University of Michigan\\
		$^\dagger$Dept. of Biomedical Engineering, University of Michigan
	}
	\vspace{1cm}
\end{center}

\input{./macro/supp}

This supplement 
elaborates upon methodology details
and presents additional results
that were excluded 
from the main body
of the manuscript \cite{nataraj::fpm}
due to word limits.
\S\ref{s,est}
details our implementations
of PERK and three other estimators
used in myelin water imaging experiments.
\S\ref{s,sim}
describes additional simulation studies
that investigate reasons for differences 
between the conventional and proposed myelin water imaging methods.
\S\ref{s,disc}
discusses additional advantages 
demonstrated by these extended simulations.

\section{Parameter Estimation Implementation Details}
\label{s,est}

\subsection{DESS $\ff$ Estimation}
\label{s,est,dess}

We used data arising 
from the fast SS scan combination 
described in Table~\ref{tab:acq}.
Since this scan combination
consisted of three DESS scans
and each DESS scan 
generates two signals per excitation,
this acquisition yielded $D \gets 6$ datasets.
We assumed non-exchanging two-compartment DESS signal models
\eqref{eq:model,dess-ref}-\eqref{eq:model,dess-def}
and took the same assumptions
as in Subsection~\ref{m,meth,acq}
to reduce model dependencies
to $L \gets 6$ latent parameters 
$\bmx \gets \brac{\ff,\tf{1},\tf{2},\ts{1},\ts{2},c}\tpose$
and $K \gets 1$ known parameter 
$\bmnu \gets \stx$ 
per voxel.
We focused on precisely estimating $\ff$ in WM 
from this fast DESS acquisition.
We considered the other five latent parameters
to be nuisance parameters
and thus did not evaluate the performance 
of their (possibly imprecise) estimators.

In all experiments
discussed in the main body,
we estimated $\ff$
using approximate PERK estimator \eqref{eq:perk,approx}.
We assumed a separable prior distribution
$\dist{\bmx,\bmnu} \gets 
	\dist{\ff}\dist{\tf{1}}\dist{\tf{2}}
	\dist{\ts{1}}\dist{\ts{2}}\dist{c}\dist{\stx}$.
We set fast-relaxing fraction marginal distribution 
$\dist{\ff} \gets \unif{-0.1,0.4}$ 
and deliberately sample negative values
\footnote{%
	Our two-compartment signal models
	are linear in $\ff$
	and are therefore well-defined
	for zero or even negative $\ff$ values.
}
with nonzero probability
to reduce $\ff$ estimation bias,
especially in low-$\ff$ regions.
We chose relaxation parameter marginal distributions
$\dist{\tf{1}} \gets \logunif{50,700}$ms,
$\dist{\tf{2}} \gets \logunif{5,50}$ms,
$\dist{\ts{1}} \gets \logunif{700,2000}$ms,
$\dist{\ts{2}} \gets \logunif{50,300}$ms
similar to those 
used for scan design
but with finite support.
To match the scaling 
of training and testing data,
we set $\dist{c} \gets \unif{2.2 \times 10^{-16},u}$,
with $u$ set as $10\times$ the maximum value 
of magnitude test data.
We estimated flip angle scaling marginal $\dist{\stx}$
from known $\stx$ map voxels
via kernel density estimation 
(implemented using the built-in \matlab function \texttt{fitdist}
with default options)
and then clipped the support of $\dist{\stx}$
to assign nonzero probability
only within $\brac{0.5,2}$.
We assumed noise covariance $\bmSig$
of form $\sigma^2 \eye{6}$ 
and estimated the (spatially invariant) noise variance $\sigma^2$
from Rayleigh-distributed regions
of magnitude test data,
using estimators described 
in \citep{siddiqui:64:sif}.
We sampled $N \gets 10^6$ latent and known parameter realizations
from these distributions
and evaluated two-compartment DESS signal models
\eqref{eq:model,dess-ref}-\eqref{eq:model,dess-def}
to generate corresponding complex noiseless measurements.
After adding complex Gaussian noise realizations,
we concatenated the (Rician) magnitude 
of these noisy measurements
with known parameter realizations
to construct pure-real regressors.
We used Gaussian kernel \eqref{eq:perk,kernel}
with bandwidth matrix $\bmL$ selected as 
$\bmL \gets \lambda \diag{\brac{\bmmymag\tpose, \bmmnu\tpose}\tpose}$,
where $\bmmymag \in \reals{D}$ 
and $\bmmnu \in \reals{K}$ 
denote sample means across voxels
of magnitude test image data and known parameters,
respectively.
We separately selected 
and then held fixed 
bandwidth scaling parameter $\lambda \gets 2^{0.3}$ 
and regularization parameter $\rho \gets 2^{-19}$ 
via the holdout procedure
described in \citep{nataraj:18:dfm}.
We implemented a $\paren{Z \gets 10^3}$-dimensional 
approximate feature map $\bmz$.
For training,
we used $\bmz$ to nonlinearly lift regressors into features
and then stored $\bmmx$, $\bmCxz\inv{\bmCzz + \rho\eye{Z}}$, and $\bmmz$.
For testing,
we evaluated \eqref{eq:perk,approx}
on test image data 
and the known transmit field map
on a voxel-by-voxel basis.

In the extended simulation study 
discussed in \S\ref{s,sim},
we compared PERK $\ff$ estimates
to maximum likelihood (ML) $\ff$ estimates
achieved via the variable projection method (VPM) \citep{golub:03:snl}
and grid search.
Following \citep{nataraj:17:oms,nataraj:18:dfm},
we clustered known flip angle scaling map voxels
into $20$ clusters
via $k$-means$++$ \citep{arthur:07:kmt}
and used each of the cluster means
to compute $20$ dictionaries.
Each of these dictionaries consisted
of nearly $8 \times 10^6$ signal vectors
computed using finely spaced samples
on an unrealistically narrow feasible region
consisting of a hypercube
with boundaries set 
as $\brac{-0.1,0.4}$ in $\ff$ 
and $\pm 20$\% away from the truth
in other latent parameter dimensions.
Iterating over clusters,
we generated each cluster's dictionary
and applied VPM and grid search
over magnitude image data voxels
assigned to that cluster.

\subsection{MESE $\mwf$ Estimation}
\label{s,est,mese}

We compared $\ff$ estimates
from our optimized DESS acquisition
to two conventional MWF $\mwf$ estimates
from a MESE acquisition.
The first conventional MWF estimate \citep{mackay:94:ivv}
is related to the solution
of a nonnegative least-squares (NNLS) problem \citep{lawson:74}
\begin{align}
	\esta{\bmx}{\bmy} \in \set{\argmin{\bmx \in \setX} \norm{\bmy - \bmA \bmx}_2^2},
	\label{eq:meth,nnls}
\end{align}
where $\bmy \in \reals{D}$ here denotes 
MESE echo amplitudes at $D$ echo times;
$\bmA \in \reals{D \times L}$ models the $D$ MESE signals
as weighted sums of $L$ discrete $\Tt$ component signals;
$\setX \subset \reals{L}$ is the nonnegative orthant;
and
$\bmx \in \setX$ here denotes $L$ component weights.
Whereas solutions to \eqref{eq:meth,nnls} tend to be sparse
for $L>D$ as is typical,
researchers have suggested
that spectral distributions are more likely smooth
in biological tissue \citep{kroeker:86:aob}.
For smoother \invivo spectrum estimates 
and for improved problem conditioning,
we also compared 
to a second MWF estimate \citep{whittall:89:qio}
that is related to the solution
of a regularized NNLS problem (RNNLS)
\begin{align}
	\esta{\bmx}{\bmy} \in \set{\argmin{\bmx\in\setX} \norm{\bmy-\bmA\bmx}_2^2
		+ \beta \norm{\bmx}_2^2},
	\label{eq:meth,rnnls}
\end{align}
where $\beta>0$ is a regularization parameter.
Similar to \citep{mackay:94:ivv} or \citep{whittall:89:qio},
each of the two conventional MWF estimators 
are then respectively defined as 
$\mwfest := \paren{\bmi\tpose \esta{\bmx}{\cdot}}/\norm{\esta{\bmx}{\cdot}}_1$,
where $\esta{\bmx}{\cdot}$ is given
by \eqref{eq:meth,nnls} or \eqref{eq:meth,rnnls}
and $\bmi \in \set{0,1}^L$ is in both cases nonzero
only in entries corresponding to $\Tt \in \brac{15,40}$ms.
As recommended in recent MESE MWF imaging literature \citep{prasloski:12:aos},
we computed MESE signal amplitudes
using the extended phase graph (EPG) formalism \citep{hennig:88:mis}
to account for stimulated echo signal contributions
that persist in MESE due to non-ideal refocusing pulses.
Note that the conventional EPG-based MESE model ignores exchange.
We accounted for non-ideal refocusing
by assuming $\stx$ is known. 
We also accounted for incomplete recovery
by assuming bulk $\To$ is known. 
To circumvent separate EPG simulations for every voxel,
we clustered known $\stx,\To$ map voxels
into $100$ clusters via $k$-means$++$ \citep{arthur:07:kmt}
and ran $100$ EPG simulations 
using each of the cluster means.
Iterating over clusters,
we generated each cluster's system matrix 
and solved \eqref{eq:meth,nnls} and \eqref{eq:meth,rnnls} 
for MESE image voxels assigned to that cluster.
We constructed NNLS and RNNLS MWF estimates
by estimating $L \gets 100$ spectral component images
from $D \gets 32$ MESE measurement images
(reducing $L$ did not appreciably influence results).
We manually fixed RNNLS regularization parameter $\beta \gets 2^{-13}$
as a small value that provided reasonable \invivo results.
We solved \eqref{eq:meth,nnls} and \eqref{eq:meth,rnnls}
using the \matlab function \texttt{lsqnonneg} 
with default options.

\section{Extensions to Simulation Studies}
\label{s,sim}

This section describes additional simulation studies
that aid in understanding reasons
for differences in MESE $\mwf$ and DESS $\ff$ estimates.
To help assess whether differences can be explained
by the unconventional PERK $\ff$ estimator,
\S\ref{s,sim,2comp} extends the two-compartment simulation
described in the main body
by adding conventional ML $\ff$ estimation.
To help assess whether differences can be explained
by the idealized two-compartment DESS signal model,
\S\ref{s,sim,3comp} investigates estimator performance
when voxel data is simulated
to more realistically arise from three water compartments.
	
\subsection{Extension to Two-Compartment Simulation}
\label{s,sim,2comp}

We extended the two-compartment simulation study
described in \S\ref{m,meth,sim}
to now include ML $\ff$ estimates
(these results were omitted in the main body
for consistency across experiments).
We estimated $\ff$ from the same noisy magnitude DESS images
and the same known $\kappa$ maps
as in the main body,
now using the dictionary-based grid search ML estimator 
described in \S\ref{s,est,dess}
in addition to the PERK estimator as before.
ML estimation took $17726$s (nearly $5$h). 
As mentioned in the main body,
PERK training and testing 
respectively took $33.8$s and $1.0$s.
	
\begin{figure}[!t]
	\centering
	\subfigure{%
		\includegraphics [width=\textwidth] {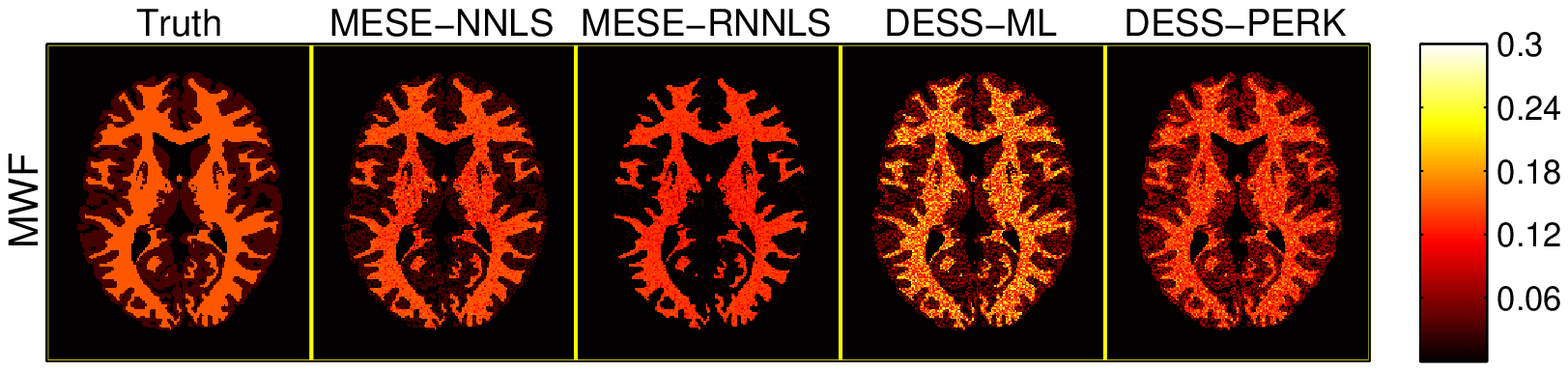}
		\label{fig:2comp-ext,im}
	}
	\hspace{0cm}
	\subfigure{%
		\includegraphics [width=\textwidth,trim=0 0 0 25,clip] {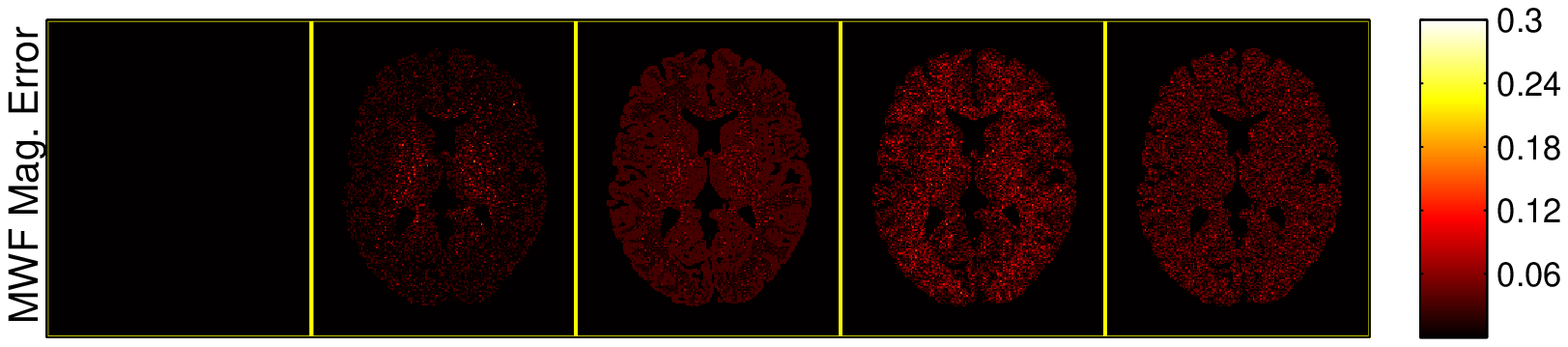}
		\label{fig:2comp-ext,err}
	}
	\caption{%
		NNLS/RNNLS MESE $\mwf$ and ML/PERK DESS $\ff$ estimates 
		alongside corresponding magnitude error images,
		in a two-compartment simulation
		where none of the associated estimators
		incur bias due to model mismatch.
		Voxels not assigned WM- or GM-like compartmental fractions
		are masked out in post-processing for display.
		Table~\ref{tab:2comp-ext} presents corresponding sample statistics.
	}
	\label{fig:2comp-ext}
\end{figure}

Fig.~\ref{fig:2comp-ext} extends Fig.~\ref{fig:2comp}
by adding the resulting DESS ML $\ff$ estimate
alongside a corresponding magnitude difference image
with respect to the ground truth $\mwf \equiv \ff$ map.
The PERK $\ff$ estimate visibly exhibits less error 
in WM-like voxels
than the ML $\ff$ estimate,
perhaps in part because PERK tuning parameters $\paren{\lambda,\rho}$
were optimized via holdout 
for estimating WM-like $\ff$ values.
Both the ML and PERK $\ff$ estimates 
exhibit less spatial variation in error maps
than MESE $\mwf$ estimates,
suggesting that reduced transmit field sensitivity
is not a property of the PERK estimator,
but rather due to consideration
of transmit field variation 
during acquisition design.
  
\begin{table}[!t]
	\centering
	\begin{tabular}{r | r r}
		\hline
		\hline
													& WM 															& GM \\
		\hline
		True $\mwf\equiv\ff$	& $0.15$ 													& $0.03$ \\
		\hline
		MESE-NNLS $\mwfest$ 	&	\mnstd{0.1375}{0.0187} (0.0225) & \mnstd{0.0203}{0.01296} (0.0162) \\
		MESE-RNNLS $\mwfest$ 	& \mnstd{0.1285}{0.0146} (0.0260) & \mnstd{0.00207}{0.00524} (0.02841) \\
		\hline
		DESS-ML $\ffest$ 			& \mnstd{0.1590}{0.0433} (0.0442) & \mnstd{0.0334}{0.0272} (0.0274) \\
		DESS-PERK $\ffest$ 		& \mnstd{0.1352}{0.0267} (0.0305) & \mnstd{0.0436}{0.0267} (0.0299) \\
		\hline
		\hline
	\end{tabular}
	\caption{%
		Sample means $\pm$ sample standard deviations (RMSEs)
		of NNLS/RNNLS MESE $\mwf$ estimates
		and ML/PERK DESS $\ff$ estimates
		in a two-compartment simulation
		where none of the associated estimators
		incur bias due to model mismatch.
		Sample statistics are computed 
		over $7810$ WM-like and $9162$ GM-like voxels.
		Each sample statistic is rounded off 
		to the highest place value
		of its (unreported) standard error,
		computed via formulas in \citep{ahn:03:seo}.
		Fig.~\ref{fig:2comp-ext} presents corresponding images.
	}
	\label{tab:2comp-ext}
\end{table}

Table~\ref{tab:2comp-ext} extends Table~\ref{tab:2comp}
by adding ML $\ff$ sample statistics.
The PERK $\ff$ estimate
is more precise but less accurate
than the ML $\ff$ estimate 
because it is a Bayesian estimator \cite{nataraj:18:dfm}.
PERK $\ff$ estimates exhibit better WM RMSE
and slightly worse GM RMSE
than ML $\ff$ estimates.
This extended simulation suggests
that PERK is at least as suitable as grid search
for reasonable $\ff$ estimation in WM.

\subsection{Three-Compartment Simulation with Model Mismatch}
\label{s,sim,3comp}

We next simulated data
to arise from three non-exchanging water compartments
with myelin water-like $\paren{500,20}$ms,
cellular water-like $\paren{1000,80}$ms,
and 
free water-like $\paren{3000,3000}$ms
(longitudinal, transverse) relaxation time constants
selected based on \citep{mackay:94:ivv,deoni:11:com}.
With this three-compartment ground truth,
the aforementioned MESE MWF estimators 
could incur bias due to their bulk-$\To$ assumption
and the aforementioned DESS fast-fraction estimators 
could incur bias due to their two-compartment assumption.
Thus $\mwf$ and $\ff$ are not equivalent here
and their estimates need not necessarily be comparable.
We assigned (myelin, cellular, free) water-like fractions
of $\paren{0.15,0.82,0.03}$ in WM,
and $\paren{0.03,0.94,0.03}$ in GM.
We simulated data 
otherwise exactly as detailed in \S\ref{m,meth,sim}
to yield MESE image datasets
with SNR ranging from 24-795 in WM
and 29-862 in GM
and to yield DESS image datasets
with SNR ranging from 24-221 in WM
and 30-241 in GM,
where SNR is computed via \eqref{eq:meth,snr}.
We estimated $\mwf$ 
from noisy magnitude MESE images
and known bulk $\To$ and $\stx$ maps
by solving NNLS \eqref{eq:meth,nnls}
and RNNLS \eqref{eq:meth,rnnls} problems
as explained in \S\ref{s,est,mese}.
We estimated $\ff$ 
from noisy magnitude DESS images and known $\stx$ maps
using ML and PERK estimators,
as explained in \S\ref{s,est,dess}.
NNLS and RNNLS respectively took $42.7$s and $69.2$s.
ML estimation took $17681$s (nearly 5h),
while PERK training and testing respectively took $34.2$s and $1.1$s.

\begin{figure}[!t]
	\centering
	\subfigure{%
		\includegraphics [width=\textwidth] {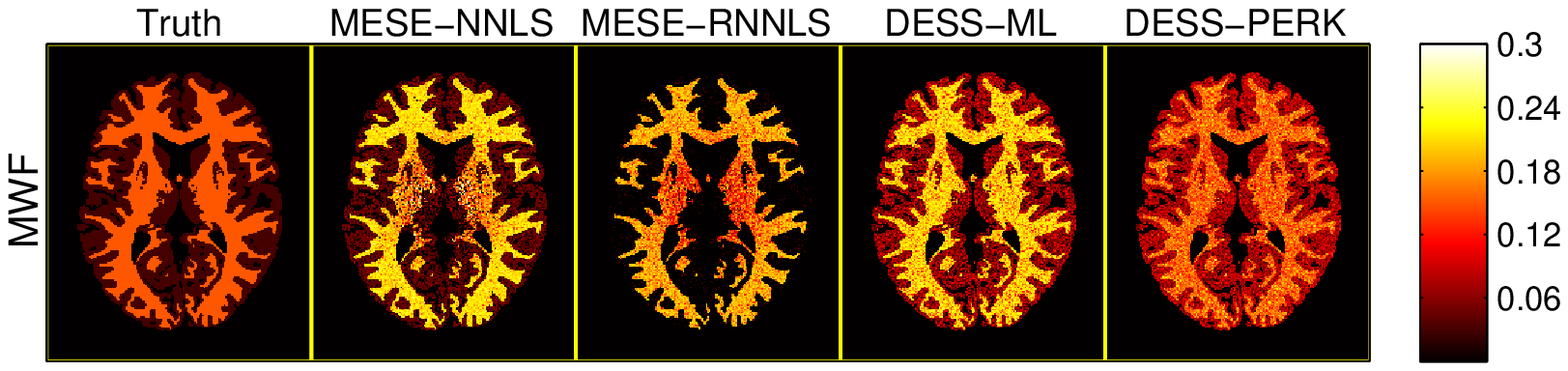}
		\label{fig:3comp-ext,im}
	}
	\hspace{0cm}
	\subfigure{%
		\includegraphics [width=\textwidth,trim=0 0 0 25,clip] {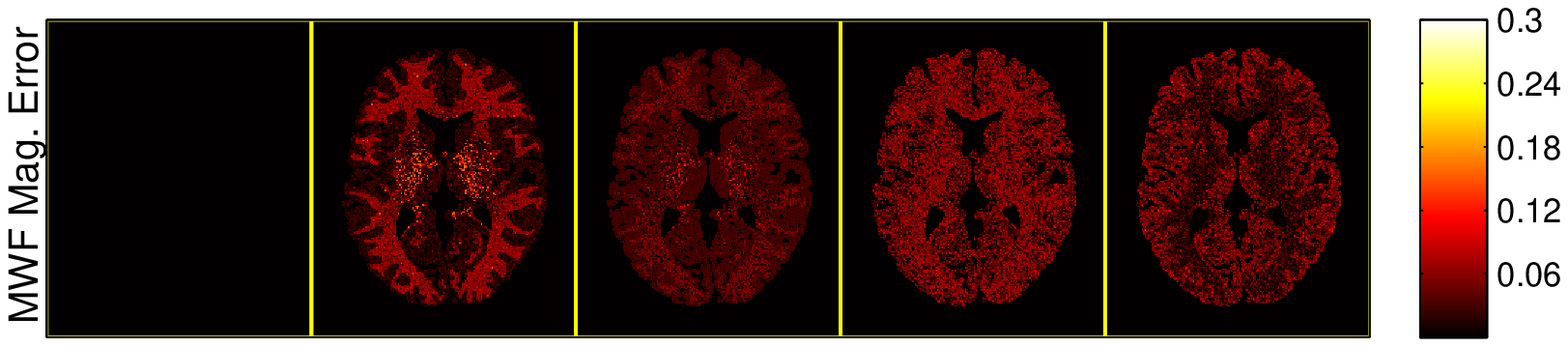}
		\label{fig:3comp-ext,err}
	}
	\caption{%
		NNLS/RNNLS MESE $\mwf$ and ML/PERK DESS $\ff$ estimates 
		alongside corresponding magnitude error images,
		in a three-compartment simulation
		where any of the associated estimators
		could incur bias due to model mismatch.
		Voxels not assigned WM- or GM-like compartmental fractions
		are masked out in post-processing for display.
		Table~\ref{tab:3comp-ext} presents corresponding sample statistics.
	}
	\label{fig:3comp-ext}
\end{figure}

Fig.~\ref{fig:3comp-ext} compares NNLS and RNNLS $\mwf$ estimates
as well as ML and PERK $\ff$ estimates
alongside magnitude difference images
with respect to the ground truth MWF.
The PERK $\ff$ estimator achieves the lowest errors in WM
but overestimates in GM
(as does the ML $\ff$ estimator),
causing reduced WM/GM contrast
relative to other estimators.
Unlike both $\ff$ estimates,
both $\mwf$ estimates visibly exhibit systematic error
due to flip angle spatial variation,
despite perfect knowledge of $\stx$.
All estimates are higher
(though to varying degrees)
than corresponding estimates 
presented in Fig.~\ref{fig:2comp-ext},
indicating some sensitivity to model mismatch.
Except for PERK $\ff$ estimates in WM
and RNNLS $\mwf$ estimates in GM,
all estimates exhibit greater error
than corresponding estimates 
presented in Fig.~\ref{fig:2comp-ext},
indicating that in most cases 
model mismatch is detrimental to estimation performance.

\begin{table}[!t]
	\centering
	\begin{tabular}{r | r r}
		\hline
		\hline
													& WM 															& GM \\
		\hline
		True $\mwf\equiv\ff$	& $0.15$ 													& $0.03$ \\
		\hline
		MESE-NNLS $\mwfest$ 	&	\mnstd{0.1910}{0.0463} (0.0618) & \mnstd{0.0349}{0.0192} (0.0198) \\
		MESE-RNNLS $\mwfest$ 	& \mnstd{0.1699}{0.0354} (0.0406) & \mnstd{0.00272}{0.00673} (0.02809) \\
		\hline
		DESS-ML $\ffest$ 			& \mnstd{0.1987}{0.0275} (0.0559) & \mnstd{0.0632}{0.0280} (0.0434) \\
		DESS-PERK $\ffest$ 		& \mnstd{0.1576}{0.0243} (0.0254) & \mnstd{0.0754}{0.0231} (0.0510) \\
		\hline
		\hline
	\end{tabular}
	\caption{%
		Sample means $\pm$ sample standard deviations (RMSEs)
		of NNLS/RNNLS MESE $\mwf$ estimates
		and ML/PERK DESS $\ff$ estimates
		in a three-compartment simulation
		where any of the associated estimators
		could incur bias due to model mismatch.
		Sample statistics are computed 
		over $7810$ WM-like and $9162$ GM-like voxels.
		Each sample statistic is rounded off 
		to the highest place value
		of its (unreported) standard error,
		computed via formulas in \citep{ahn:03:seo}.
		Fig.~\ref{fig:3comp-ext} presents corresponding images.
	}
	\label{tab:3comp-ext}
\end{table}

Table~\ref{tab:3comp-ext} compares sample statistics
of NNLS and RNNLS $\mwf$ estimates
as well as ML and PERK $\ff$ estimates,
computed over the same WM-like and GM-like ROIs
as in Table~\ref{tab:2comp-ext}.
Several estimates now differ from true values
by more than one standard deviation,
indicating significant bias due to model mismatch
in these cases.
The PERK $\ff$ estimator is most accurate 
and achieves the lowest RMSE in WM,
but also suffers from the highest RMSE in GM.
The NNLS $\mwf$ estimator is most accurate
and achieves the lowest RMSE in GM,
but also suffers from the highest RMSE in WM.
RNNLS $\mwf$ (PERK $\ff$) estimates are now
both more accurate and more precise
than NNLS $\mwf$ (ML $\ff$) estimates
in WM,
suggesting that regularization may be beneficial
in cases of model mismatch.
Perhaps surprisingly,
RNNLS $\mwf$ and PERK $\ff$ estimates 
do not differ significantly in WM
(but do differ in GM)
suggesting that these WM estimates may be comparable
even when characterizing 3-compartment systems,
at least for the nominal ground-truth values selected here.

\section{Further Discussion}
\label{s,disc}

Extended simulations provide evidence
that PERK is well-suited for DESS $\ff$ estimation.
Idealized two-compartment simulations demonstrate 
that PERK and standard ML $\ff$ estimators
achieve comparable RMSE in WM- and GM-like voxels,
but PERK is more than $500\times$ faster.
More realistic three-compartment simulations reveal
that conventional MESE $\mwf$ estimates are sensitive
to unaccounted variable $\To$-recovery rates across compartments
and accounted flip angle spatial variation
while DESS $\ff$ estimates are sensitive 
to relaxation in an unaccounted third compartment,
though the DESS PERK $\ff$ estimator is most robust in WM
to these sources of model misspecification errors.

Taken together with the results in \cite{nataraj:18:dfm},
results herein also provide evidence
that the PERK estimator 
can maintain good performance
while scaling more gracefully
with the number of unknowns per voxel $L$
than conventional ML estimators.
In an application
with $L\gets3$ unknowns \cite{nataraj:18:dfm},
PERK was consistently at least $140\times$ faster
than two well-suited ML estimators
and achieved comparable performance 
in simulation, phantom, and \invivo studies.
In myelin water imaging simulations ($L\gets6$),
PERK was consistently at least $500\times$ faster
than an ML estimator
achieved via unrealistically narrow grid search
around the ground truth.
In early myelin water imaging \invivo experiments
on other precision-optimized SPGR/DESS datasets
from the same healthy volunteer,
PERK took comparable time 
($\sim$1m including training)
and produced similar $\ff$ estimates
as reported here
while a more realistically-constrained grid search
took about $68$ CPU-days
(running on $24$ nodes of a computing cluster).
We omitted \invivo and \exvivo ML results here
because these early experiments
produced poor ML $\ff$ estimates,
likely due to multiple global minima 
of the associated ML cost function.
Since PERK training time scales negligibly 
with the number of voxels,
all of these acceleration factors 
would scale roughly linearly
with the number of reconstructed slices
for full-volume parameter estimation problems.

\end{document}